\documentclass[journal=acsnano,manuscript=paper]{achemso}
\SectionNumbersOn

\usepackage{chemformula} 
\usepackage[T1]{fontenc} 
\usepackage{hyperref}
\usepackage{chngcntr}

\newcommand{\be}{\begin{equation}}
\newcommand{\ee}{\end{equation}}
\newcommand{\bea}{\begin{eqnarray}}
\newcommand{\eea}{\end{eqnarray}}

\usepackage{graphicx}
\usepackage{bm}
\usepackage{color}
\usepackage{romannum}
\usepackage{comment}
\usepackage[numbers]{natbib}

\author{Akshaya Kumar Jena}
\affiliation{Department of Chemistry, Birla Institute of Technology and Science Pilani, Pilani Campus, Vidya Vihar, Pilani-333031, Rajasthan, India}

\author{Aashima}
\affiliation{Department of Chemistry, Birla Institute of Technology and Science Pilani, Pilani Campus, Vidya Vihar, Pilani-333031, Rajasthan, India}

\author{Pritam Kumar Jana}
\email{pritam.jana@pilani.bits-pilani.ac.in} 
\affiliation{Department of Chemistry, Birla Institute of Technology and Science Pilani, Pilani Campus, Vidya Vihar, Pilani-333031, Rajasthan, India}
\altaffiliation{These authors contributed equally to this work.}

\author{Bortolo Matteo Mognetti}
\email{Bortolo.Matteo.Mognetti@ulb.be} 
\affiliation{ Interdisciplinary Center for Nonlinear Phenomena and Complex Systems, Université Libre de Bruxelles (ULB), B-1050 Brussels, Belgium }%
\altaffiliation{These authors contributed equally to this work.}

\title[Colloidal Layer Deposition with a Controllable Number of Layers and Compositional Order]
  {Colloidal Layer Deposition with a Controllable Number of Layers and Compositional Order}

\abbreviations{DNA, DNACC}

\begin{document}

\begin{tocentry}

\includegraphics[width=\linewidth]{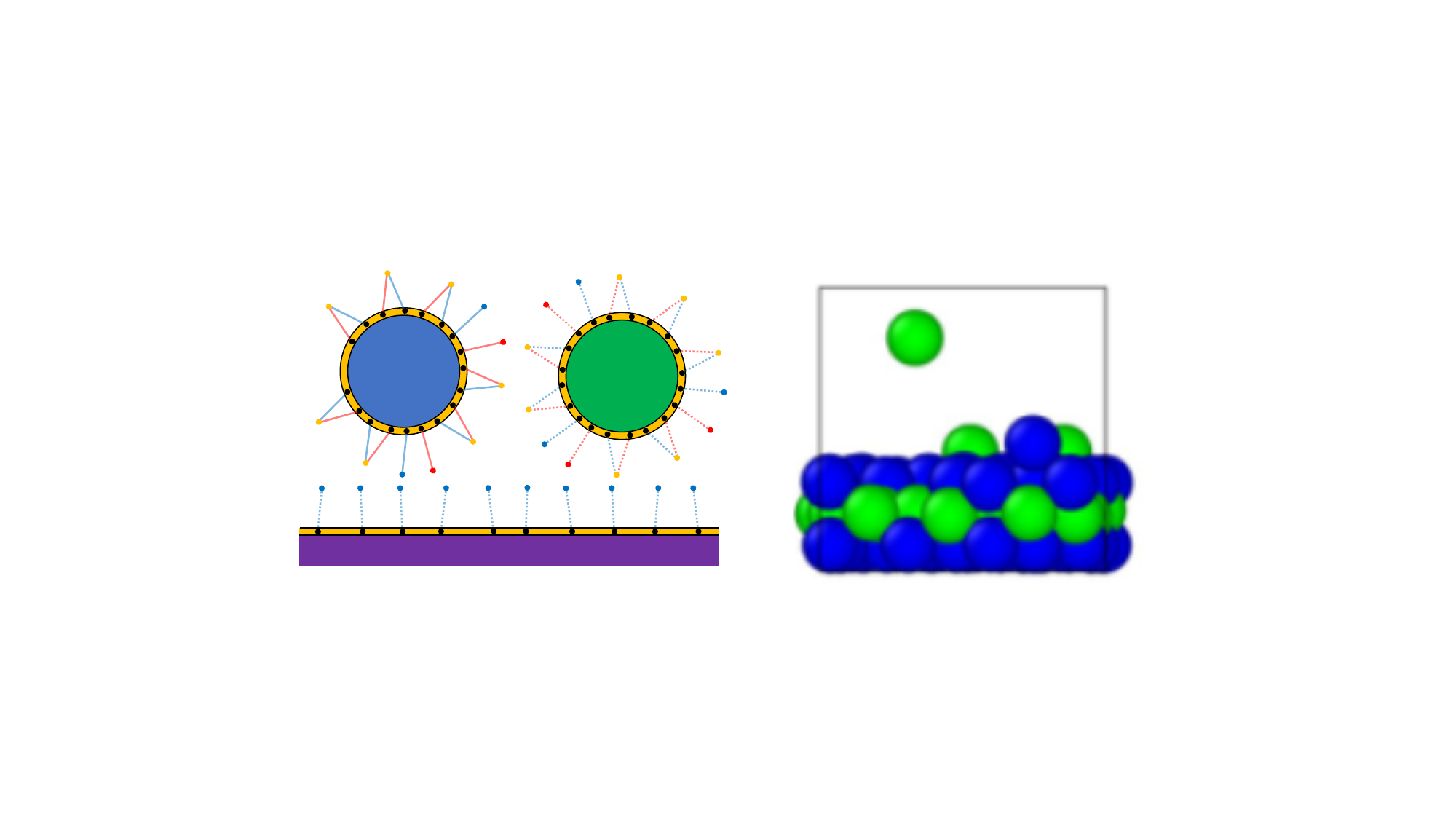}

\end{tocentry}

\begin{abstract}

We design a system with a binary suspension of colloids and a surface that triggers the self-assembly of crystallites with a finite thickness.
The proposed design allows controlling the number of layers forming the aggregate and constrains the two types of particles to lie on different planes. 
These functionalities are achieved by decorating the colloids and the surface with multiple DNA oligomers featuring specific interactions. 
Similarly to what was proposed in previous studies,
the surface triggers a chain of reactions between DNA oligomers, leading to localized self-assembly. 
System parameters control the thickness of the aggregates.
In this work, compositional order is achieved by engineering the reaction kinetics between DNA oligomers in a way that limits interactions between colloids of the same type. 
We validate our design using theory and reaction-diffusion simulation algorithms, which capture the multibody nature of the interactions. 
This work demonstrates how engineering the reaction kinetics provides a new tool for controlling the morphology of aggregates assembled by DNA.

\end{abstract}


\section{Introduction}

Over the past two decades, DNA-mediated interactions have emerged as a powerful tool for programming colloidal self-assembly.\cite{Mirkin_Nature_1996,Alivisatos_Nature_1996,Jones_Science_2015} Colloids functionalized with DNA oligomers that selectively bind complementary sequences\cite{SantaLucia_PNAS_1998,fornace2020unified} have been employed to self-assemble a wide variety of aggregates—both disordered\cite{Varrato_PNAS_2012} and crystalline\cite{Liu_Science_2016,Wang_NatComm_2017,Ducrot_NatMat_2017,Macfarlane_AngChem_52,Nykypanchuk_Nature_2008,Mirkin_Nature_2008,ye2026preparation}—with precise control over their structural and morphological properties.\cite{halverson2013dna,patra2017layer,jana2019surface}

However, the self-assembly potential of DNA-functionalized systems is often limited by slow kinetics. In multivalent systems, particle-particle interactions are highly sensitive to control parameters such as temperature and coating densities. These factors can lead to overly sticky interactions,\cite{marbach2022nanocaterpillar,lowensohn2022sliding,dai2026phase} which impedes the relaxation of the system toward the thermodynamically stable state. Researchers have tackled this issue by optimizing design parameters (e.g., increasing coating densities\cite{wang2015crystallization}) and employing annealing protocols.\cite{biancaniello2005colloidal,zhou2020programming,hensley2023macroscopic} Advanced sequence designs based on the toehold exchange mechanism of Zhang and Winfree \cite{zhang2009control} further address this challenge by accelerating binding and unbinding reactions between complementary sticky ends, thus speeding up thermalization.\cite{parolini2016controlling,rogers-manoharan}

In this manuscript, we demonstrate that the toehold exchange mechanism can also enhance the degree of control over self-assembled aggregates in multicomponent systems.
 Engineering the reaction kinetics between sticky ends allows accelerating specific particle-particle interactions while suppressing others. This selective control constrains the self-assembly pathway, directing the system toward desired outcomes. Using the systems developed by Jana et al. \cite{jana2019surface,jana2020self} and Lanfranco et al. \cite{lanfranco2020adaptable}, we explore a setup where a surface decorated with DNA oligomers in a bath of self-protected colloids (colloids carrying DNA sticky ends forming intra-particle loops) initiates cascade reactions. These reactions lead to the self-assembly of finite-sized aggregates (see Fig.~\ref{Fig:Design}).\cite{jana2019surface,lanfranco2020adaptable} 
 Refs.~\cite{jana2019surface,lanfranco2020adaptable} clarified how the aggregate thickness is controlled by the number of DNA oligomers {\em per} particle and the chemical potential of the colloids. In this contribution, we extend the design of Refs.~\cite{jana2019surface,lanfranco2020adaptable} by considering a binary system and employing a toehold-exchange mechanism\cite{zhang2009control} that favors interactions between distinct types of colloids.
This approach results in the self-assembly of finite-sized aggregates with compositional order, including crystals made of alternating stacks of same-type particles. Such types of morphology cannot be yielded using thermodynamic designs as intra-particle loops implies the possibility of forming inter-particle bridges (which mediate self--assembly).
 The proposed strategy to limit interaction between particles of the same type is general and can be used in responsive systems employing self-protected colloids.

DNA is not only a versatile "glue" but also a medium for information processing and responsive behavior.\cite{schulman2012robust,barish2009information,Jones_Science_2015,evans2024pattern} This  feature has been extensively harnessed in DNA nanotechnology for applications such as sensing,\cite{Thaxton_PNAS_2009,Chapman2015} targeting, and delivery.\cite{martinez2011designing,licata2008kinetic,he2023self,hong2025advances} Despite these advances, complex responsive interactions remain underexplored in self-assembly designs, especially in DNA-functionalized colloid systems. Our work provides a novel example of complex information processing in self-assembly, opening new avenues for designing responsive, programmable, and information-processing colloidal systems.

\section{Methods}

\subsection{The Model}

\begin{figure}[h]
\includegraphics[scale=0.45]{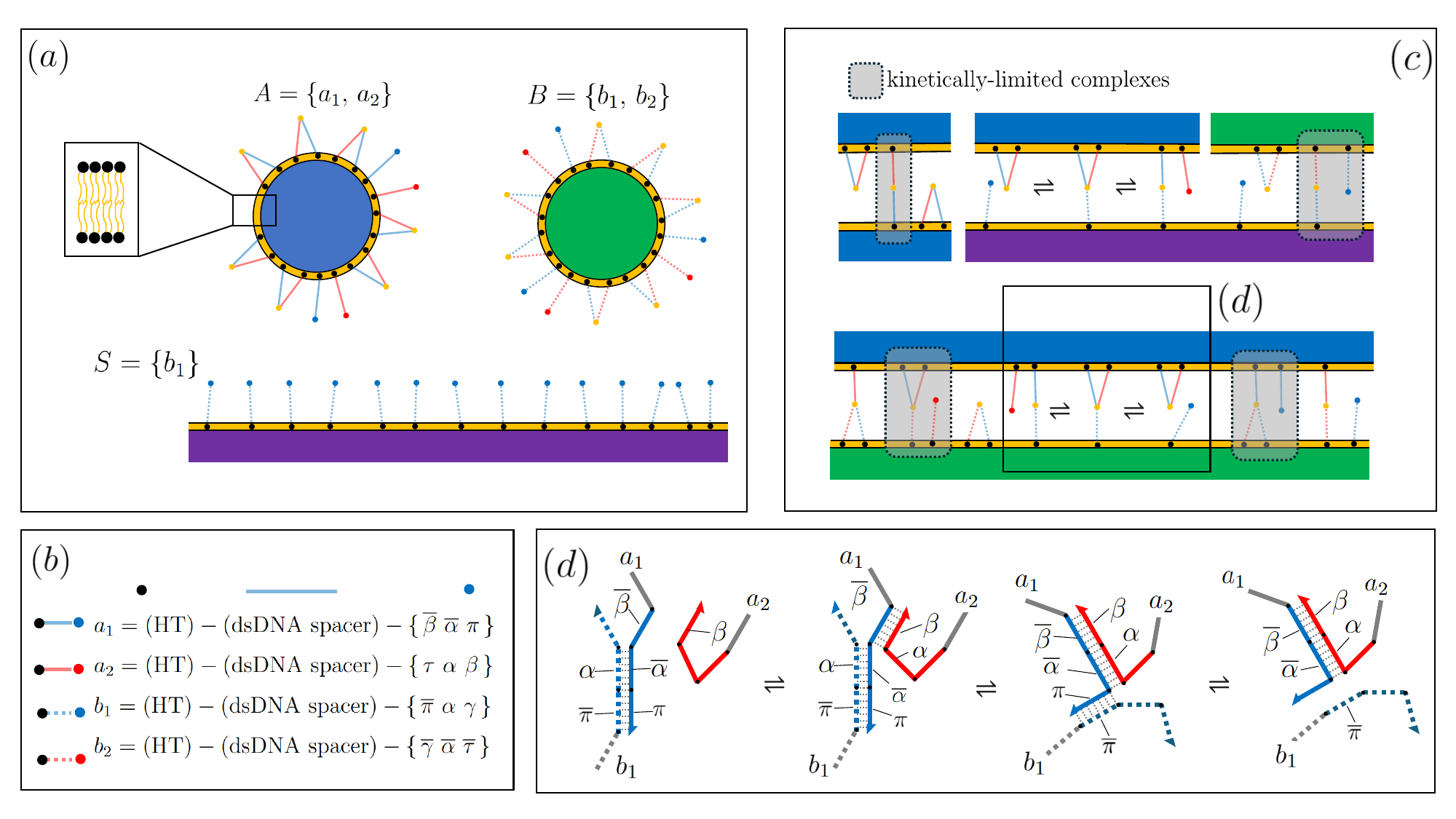} 
\caption{{\bf The model.} 
$(a)$ Illustration of a binary system comprising colloids (A and B) functionalized with two pairs of ligands ($a_1$, $a_2$ and $b_1$, $b_2$) in the presence of a surface functionalized with a single type of ligand ($b_1$). 
$(b)$ Schematic representation of ligands, consisting of a hydrophobic tag (anchoring the DNA to supported lipid bilayers), a rigid double-stranded DNA spacer, and a sticky end composed of three modules.
$(c)$ List of all possible complexes in the system, including loops and bridges that cross-link particles and link particles to the surface. Complexes that can be kinetically suppressed are highlighted in gray (see SI Fig.~S1). For clarity, particles are represented as planes. $(d)$ Molecular rendering of the toehold-exchange mechanism of Zhang and Winfree,\cite{zhang2009control} enabling transitions between intra-particle loops and inter-particle bridges without denaturation of the $\alpha$ module.}
\label{Fig:Design}
\end{figure}

Fig.~\ref{Fig:Design} presents our system. We consider rigid particles carrying mobile ligands composed of rigid rods of double-stranded (ds) DNA of length $L$, tipped by sticky ends made of single-stranded (ss) DNA oligomers (Fig.~\ref{Fig:Design}$a$). This class of building blocks can be fabricated using colloid-supported lipid bilayers\cite{Meulen_JACS_2013,rinaldin2018colloidal} functionalized with DNA linkers anchored to the membrane via hydrophobic tags (HT) (Fig.~\ref{Fig:Design}$a$ and $b$).\cite{Beales_JPCA_2007,Hadorn_PNAS_2012,Parolini_NatComm_2015} Particles functionalized with mobile ligands have been employed to engineer multi-body interactions.\cite{Angioletti-Uberti_PRL_2014} For further details on the design and self-assembly capabilities of this class of materials, we refer to a literature review.\cite{mognetti2019programmable}

We study a binary system comprising two types of colloids (A and B), each carrying two types of sticky ends ($a_1$, $a_2$ and $b_1$, $b_2$) along with a surface decorated with $b_1$-type ligands (Fig.~\ref{Fig:Design}$a$). $N_L$ is the number of ligands of a given type {\em per} particle.
The sequence design of the DNA sticky ends (Fig.~\ref{Fig:Design}$b$) enables the formation of specific two-strand complexes: $a_1 a_2$, $b_1 b_2$, $a_1 b_1$, and $a_2 b_2$. For instance, $a_1 a_2$ represents a complex formed by the hybridization of sticky ends $a_1$ and $a_2$. Two-strand complexes lead to the formation of intra-particle loops ($a_1 a_2$ and $b_1 b_2$), and inter-particle bridges between particles of the same and different type ($a_1 a_2$, $b_1 b_2$, and $a_1 b_1$, $a_2 b_2$, respectively). All possible complexes are shown in Fig.~\ref{Fig:Design}$c$.

A class of sequence designs resulting in these pairings is illustrated in Fig.~\ref{Fig:Design}$b$. Each sticky end comprises three modules (subgroups of bases) denoted by Greek letters (e.g., $\alpha$). Complementary sequences are represented with a bar (e.g., if $\alpha=3'-CTGGA-5'$ then $\bar \alpha =3'-TCCAG-5'$). Each sticky end consists of a central module ($\alpha$ or $\bar \alpha$) flanked by two side modules properly chosen from four possible sequences ($\pi$, $\tau$, $\gamma$, $\beta$) or their complements. All modules are not palindromic. Two-strand complexes form through hybridization of the central module and one of the side modules. Additionally, the sequences in Fig.~\ref{Fig:Design}$b$ allow the formation of the following three-strand complexes: $a_1 a_2 b_1$, $a_1 a_2 b_2$, $b_1 b_2 a_1$, and $b_1 b_2 a_2$. In these three-strand complexes (e.g., $a_1 a_2 b_2$, see Fig.~\ref{Fig:Design}$d$), two strands ($a_1$ and $b_2$) compete to bind the central module of the third strand ($a_2$) while binding its side modules  ($\beta$ and $\tau$). This structure facilitates the transition between two two-strand complexes ($a_1 a_2$ and $a_2 b_2$) by forming a three-strand complex.\cite{zhang2009control} In this process, $b_2$ binds the $\tau$ module (toehold module\cite{zhang2009control}) of $a_2$, progressively displacing $a_1$ from $a_2$ and facilitating its detachment. This displacement mechanism enables transitions between intra-particle loops and inter-particle bridges, thereby accelerating self-assembly dynamics, as demonstrated by Parolini {\em et al.}~\cite{parolini2016controlling}.

We do not specify the sequence details further. However, we emphasize the importance of minimizing unwanted interactions between modules. This can be easily achieved given the availability of tens of orthogonal sequences. The sticky ends' DNA sequences are characterized by their hybridization free energy when free in solution\cite{SantaLucia_PNAS_1998,fornace2020unified} (i.e.~not tethered to the surface of the particles). For simplicity, we assume that the hybridization free energy\cite{SantaLucia_PNAS_1998,fornace2020unified} of all two- and three-strand complexes (measured using the reference concentration $\rho_0$) are equal to $\Delta G_0$ and $\Delta G_0+\Delta G_T$, respectively. $\Delta G_0-\Delta G_T$ roughly corresponds to the hybridization free energy between $\alpha$ and $\overline \alpha$ while $\Delta G_T$ to the hybridization free energy between two complementary toehold modules (Fig.~\ref{Fig:Design}$b$, $d$). Finally, the sequence design in Fig.~\ref{Fig:Design}$d$ permits the formation of four-strand complexes ($a_1a_2b_1b_2$), but these are neglected in the following.  Four-strand complexes can be suppressed by enhancing the free-energy barriers due to steric interactions between non-complementary DNA segments, which limit the hybridization of complementary modules \cite{srinivas2013biophysics}. Steric interactions increase with the number of strands entering the complex.

\subsection{Engineering the Reaction Pathway}

In this work, we consider the strong association limit, where the pairing between the $\alpha$ and $\overline \alpha$ modules (Fig.~\ref{Fig:Design}$b$) becomes effectively irreversible, while the denaturation of paired toehold modules remains accessible to thermal fluctuations. This condition can be readily achieved by carefully tuning the temperature and lengths of the respective modules.

Under these conditions, starting from dilute solutions, colloids in bulk adopt a self-protected state, maximizing the number of intra-particle loops on each particle. This configuration prevents the formation of inter-particle bridges, thereby inhibiting aggregation. Previous studies\cite{bachmann2016melting,jana2019surface,sciortino2020combinatorial} have shown that the diluted phase is thermodynamically stable (i.e., even when loops are allowed to open) for sufficiently low numbers of ligands per particle ($N_L$) or colloidal chemical potential ($\mu$). Instead, in the current design, the gas phase is expected to remain kinetically stable even under conditions where the assembled phase would be thermodynamically favorable.

The surface (coated with ligands at a density $\rho_S$), on the contrary, is not self-protected (Fig.~\ref{Fig:Design}$a$) and can trigger self-assembly. Specifically, only A-type colloids can interact with the surface by forming three-strand complexes of the type $a_1 a_2 b_1$ (Fig.~\ref{Fig:Design}$c$). This interaction leads to the formation of $b_1 a_1$ bridges, leaving free $a_2$ linkers on surface-bound A particles. These free $a_2$ linkers can selectively bind B-type colloids, triggering the self-assembly of a second layer. The process then iterates, with A-bound B-type colloids enabling the assembly of a third layer composed of A-type particles, and so on.  In SI Fig.~1, we further detail the chain of reactions concurrent with subsequent particle attachment. Importantly, bound colloids only express one type of free linkers ($a_2$ and $b_1$, respectively, for colloids of type A and B), limiting the interaction between particles of the same type. Note that a hypothetical four-strand complex between a B-type and an A-type particle could lead to the formation of a three-strand complex plus an unwanted  $a_1$ linker on particle $A$ (Fig.~1 and SI Fig.~S1). However, due to the irreversible limit and the excess of $a_2$ linkers in bound A-type particles (SI Fig.~S1), unwanted $a_1$ linkers are expected to bind $a_2$ linkers and form loops readily.

Other contributions have used strand-displacement schemes to trigger a chain of reactions and control the self-assembly pathway.\cite{Baker_SoftMatter_2013,Tison_SoftMatter_2010,rogers-manoharan,zhang2017sequential,evans2024designing,halverson2016sequential} 
 As compared to these contributions,\cite{zhang2017sequential,evans2024designing} in our scheme, particle-particle interactions are reversible, facilitating relaxation of aggregates towards the equilibrium state.

Based on this mechanism, we anticipate that the proposed design, operating in the self-protected limit, can yield adsorbed crystallites exhibiting compositional order. Specifically, the two types of colloids (A and B) are expected to segregate into distinct planes. Beyond compositional order, the design enables control over the total number of layers by modulating the chemical potential $\mu$ and the number of linkers $N_L$.\cite{jana2019surface} 
For instance, starting under conditions where the gas phase is thermodynamically unstable, the system could potentially assemble bulk crystals. However, growth may halt due to slow reaction kinetics  (see Sec.~\ref{Sec:KL}), a phenomenon observed in earlier investigations\cite{Petitzon_SoftMatt_2016} and more recently described in systems of functionalized polymers.\cite{ranganathan2020dynamic}

In Sec.~\ref{Sec:Simulation}, we present the simulation algorithm that will be used in Sec.~\ref{Sec:few_particles} and \ref{Sec:many_particles} to validate the proposed design.

\subsection{Simulation Methods}\label{Sec:Simulation}

Each step of the simulation algorithm comprises an update of the colloids' position using Brownian dynamics and an update of the population of each type of complex using reaction dynamics.\cite{gillespie1977exact,jana2019surface,Petitzon_SoftMatt_2016} We describe the two dynamics in the next two sections and in the SI document.

\subsubsection{Reaction Dynamics}\label{Sec:Sim:Reac}

As shown in Fig.~\ref{Fig:Design}$c$, our system features loops, two-strand, and three-strand bridges. We label the number of these complexes, respectively, with $n^{x,y}_{i}$, $n^{x;y}_{i;j}$, and $n^{x,y;z}_{i;j}$. We denote the type of ligands entering the complex with $x$, $y$, and $z$ while $i$ and $j$ label the particles supporting the complex. We use a semicolon to separate indices belonging to different particles. The equilibrium numbers of complexes are then given by 
\begin{eqnarray}
   \langle n^{x;y}_{i;j} \rangle &=& \langle n^x_i \rangle \langle n^y_j \rangle e^{- \beta \Delta G_0} {\Omega_{ij} \over \Omega_i \Omega_j \rho_0} 
   \nonumber
   \\
   \langle n^{x,y}_{i} \rangle &=& \langle n^x_i \rangle  \langle n^y_j \rangle e^{- \beta \Delta G_0} {1 \over \Omega_i \rho_0} 
   \nonumber
   \\
    \langle n^{x,y;z}_{i;j} \rangle &=& \langle n^x_i \rangle  \langle n^y_i \rangle  \langle n^z_j  \rangle {e^{- \beta (\Delta G_0 + \Delta G_T)}\over n_\alpha+1} {\Omega_{ij} \over (\Omega_i)^2 \Omega_j (\rho_0)^2} 
    \label{Eq:chem_eq}
\end{eqnarray}
where $n^x_i$ is the number of free linkers of type $x$ moving on particle $i$, and the average is taken at a given set of colloids' positions $\{ {\bf r}_1, \cdots, {\bf r}_N \}$ ($N$ being the number of particles). 
$\Omega_{i,j}$ and $\Omega_i$ are volumes in configurational spaces\cite{mognetti2019programmable} available, respectively,  to inter-particle complexes (2-strand and 3-strand bridges) and intra-particle complexes (loops and free linkers). For linkers modeled as thin rods (relevant to dsDNA structures) of length much smaller than the radius of the particle ($L \ll R$), $\Omega_i$ and $\Omega_{i,j}$ can be identified with the Euclidean space spanned by the end-points of the complexes (see SI Fig.~S9).\cite{Angioletti-Uberti_PRL_2014}. Further details about the configurational terms are given in SI Sec.~2. The number of two-strand and three-strand bridges between the surface and a colloid follows equations similar to Eq.~\ref{Eq:chem_eq} but with different configurational terms as specified in SI Fig.~S9$b$ and $c$. The term $n_\alpha$ is the number of bases in the $\alpha$ module (Fig.~\ref{Fig:Design}$b$). At the same time, $n_\alpha+1$ is the number of (almost isoenergetic\cite{srinivas2013biophysics}) states featured by three-strand complexes, which contribute with a combinatorial factor to the equilibrium reaction. It is important to emphasize that the primary results of this study are not contingent on the specific configurational terms employed. Rather, they are broadly applicable across various ligand structures with sticky end sequences as in Fig.~\ref{Fig:Design}$b$.

Linked to the equilibrium constants defined in Eqs.~\ref{Eq:chem_eq} are the reaction rates. Reaction rates leading to the formation/dissolution of two-strand complexes are the following (the first and second lines refer, respectively, to bridges and loops):
\begin{eqnarray}
    k^{\it on}_{x,i;y,j} = k_0 {\Omega_{ij} \over \Omega_i \Omega_j \rho_0} \, \, ,
    & \qquad &
    k^{\it off}_{x,i;y,j} = k_0 e^{\beta \Delta G_0} \, \, ,
    \nonumber \\
    k^{\it on}_{x,y,i} = {k_0 \over \Omega_i \rho_0} \, \, ,
    & \qquad &
    k^{\it off}_{x,y,i} = k_0 e^{\beta \Delta G_0} \, \, ,
\label{Eq:rates_2}
\end{eqnarray}
where $k_0$ is a reaction rate. Unless specified, in the following we set $k_0 = 1$.  In principle, $k_0$ could be computed starting from the reaction rates of the corresponding free sticky ends in solution.\cite{zhang2009control} However, such a computation would require considering the slowdown of the diffusion motion of the sticky ends due to tethering to a lipid bilayer.
Experiments studying the denaturation of dsDNA pulled using atomic force microscopy setups have shown that forces much greater than thermal fluctuations are required to alter the structure of the double helix.\cite{ho2009force} Therefore, in Eq.~\ref{Eq:rates_2}, we consider {\it off} rates that do not depend on the tension exerted on the reacted sticky ends by the DNA linkers.

There are two reaction pathways leading to the formation of three-strand bridges, either starting from a loop or a bridge (respectively, first and second lines): 
\begin{eqnarray}
    k^{{\it on},\ell}_{x,y,i;z,j} = k_0 {\Omega_{ij} \over \Omega_i \Omega_j \rho_0}
    & \qquad &
    k^{{\it off},\ell}_{x,y,i;z,j} = k_0 {e^{\beta \Delta G_T} \over n_\alpha +1}
    \nonumber \\
    k^{{\it on},b}_{x,y,i;z,j} = {k_0 \over \Omega_j \rho_0} 
    & \qquad &
    k^{{\it off},b}_{x,y,i;z,j} = k_0 {e^{\beta \Delta G_T} \over n_\alpha +1}
\label{Eq:rates_3}
\end{eqnarray}
where $b$ and $\ell$ denote, respectively, the 'bridge' and the 'loop' pathway. Eqs.~\ref{Eq:rates_3} can be adapted to the case in which complexes are anchored to the surface by adapting the configurational contributions.

The reaction algorithm starts with calculating the affinities (rates at which one type of reaction is happening) using the rate constants given in Eqs.~\ref{Eq:rates_2}, \ref{Eq:rates_3}. We then employ the Gillespie algorithm to sample the list of reactions along with the time for them to happen.\cite{gillespie1977exact} After each reaction, the affinity matrices are updated. The list of reactions is interrupted when the cumulative reaction time exceeds the simulation timestep $\Delta t$. Details about the implementation of the algorithm can be found in our previous studies.\cite{Petitzon_SoftMatt_2016,jana2019surface} The full list of affinities is reported in SI Sec.~3.

The lateral diffusion of tethered linkers is much faster than the typical self-assembly timescale.\cite{Meulen_JACS_2013,rinaldin2018colloidal,Beales_JPCA_2007,Hadorn_PNAS_2012,Parolini_NatComm_2015} Thus, we do not track the position of each ligand (as done in other studies\cite{mitra2023coarse} targeting smaller systems) but assume that they are uniformly distributed in the corresponding configurational space (SI Sec.~2).

When running equilibrium simulations, instead of using stochastic reaction dynamics, we set the numbers of complexes to their equilibrium values given by Eqs.~\ref{Eq:chem_eq}.

\subsubsection{Diffusion Dynamics}\label{Sec:Sim:Diff}

The evolution of the colloids' positions ($\{{\bf r}_i\}$) is implemented by using Brownian dynamics updates \cite{allen2017computer} as follows  
\begin{eqnarray}
    {\bf r}_i(t+\Delta t) = {\bf r}_i(t) + \beta D {\bf f}_i\Delta t + \sqrt{2 D {\bf {\cal N}_i}(0,1) \Delta t}
\end{eqnarray}
where $D$ is the diffusion constant of the colloids in dilute conditions, ${\cal N}_i(0,1)$ is a vector sampled from the standard normal distribution, $\beta$ is the inverse temperature ($\beta=1/k_B T$, with $k_B$ being the Boltzmann constant), and $\Delta t$ is the integration time step.

By adapting Eq.~49 of Ref.\cite{mognetti2019programmable} to the present system (see SI Sec.~4 for the full derivation), we write the total force acting on particle $i$ (${\bf f}_i$) as follows:
\begin{eqnarray}
{\bf f}_i &=& \sum_{j\in v(i)} n^{(b)}_{i;j} {{\bf \nabla}_{{\bf r}_i} \Omega_{ij} \over \Omega_{ij}} - \sum_{j\in v (i)} \left(n^{(\ell)}_i {{\bf \nabla}_{{\bf r}_i} e_{ij} \over \Omega_i }  + n^{(\ell)}_j {{\bf \nabla}_{{\bf r}_i} e_{ji} \over \Omega_j }  \right) \, \, ,
\label{Eq:BD_f}
\\
n^{(\ell)}_i &=&  \sum_x n^x_i+ \sum_{x,y} n^{x,y}_{i} \, \, ,
\nonumber \\
n^{(b)}_{i;j} &=& \sum_{x,y} n^{x;y}_{i;j}+\sum_{x,y,z} (n^{x,y;z}_{i,j}+n^{x,y;z}_{j,i}) \, \, .
\nonumber
\end{eqnarray}
where $v(i)$ is the list of neighboring particles—including the surface—of particle $i$. Following the notation of Sec.\ \ref{Sec:Sim:Reac}, $n^{(\ell)}_{i}$ and $n^{(b)}_{i;j}$ denote, respectively, the total number of complexes with tethering points grafted to particle $i$, and to particle $i$ and $j$. $e_{ij}$ is the configurational volume excluded to a loop or bridge on particle $i$ by the presence of particle $j$ (SI Fig.~S9 and SI Sec.~4). As clarified in SI Sec.~2, Eq.~\ref{Eq:BD_f} neglects non-specific interactions between ligands and employs averages over all possible complexes' configurations at a given $\{ {\bf r}_i \}$.\cite{Angioletti-Uberti_PRL_2014} When particle $i$ interacts with the wall $\Omega_{ij}$, $e_{ij}$, and $e_{ji}$ are replaced, respectively, by $\Omega^s_i$, $e^s_i$, and $k^s_i$ (SI Fig.~S9 and SI Sec.~4). The number of complexes is calculated using the reaction dynamics algorithm presented in the previous section or, in equilibrium conditions ($k_0 \to \infty$ in Eqs.~\ref{Eq:rates_2}, \ref{Eq:rates_3}), their averaged values (Eqs.~\ref{Eq:chem_eq}).\cite{Angioletti-Uberti_PRL_2014} The multibody nature of the interaction arises from the fact that, similar to chemical networks,\cite{maslov2007propagation} complexes belonging to a given cluster of colloids are all correlated.

Eq.~\ref{Eq:BD_f} can be easily interpreted: the first term on the right-hand side is an attractive term mediated by all (two-strand and three-strand) bridges. 
This term is extensive in the number of bridges cross-linking two particles and is modulated by an osmotic term related to the gain in configurational volume of bridges when two particles approach (SI Fig.~S9$b$ and $c$). 
In contrast, the second and third terms are entropic repulsive terms arising from the compression of free ligands and loops caused by the hard-core repulsions between colloids and complexes (SI Fig.~S9$b$ and $c$). 
As done previously, we regularize the excluded-volume repulsion between colloids using repulsive nonspecific interactions similar to the second and third terms in Eq.~\ref{Eq:BD_f} (for details, see SI Sec.~4).

\begin{figure}
    \centering
    \includegraphics[scale=0.50,trim=0cm 0cm 0cm 0cm,clip]{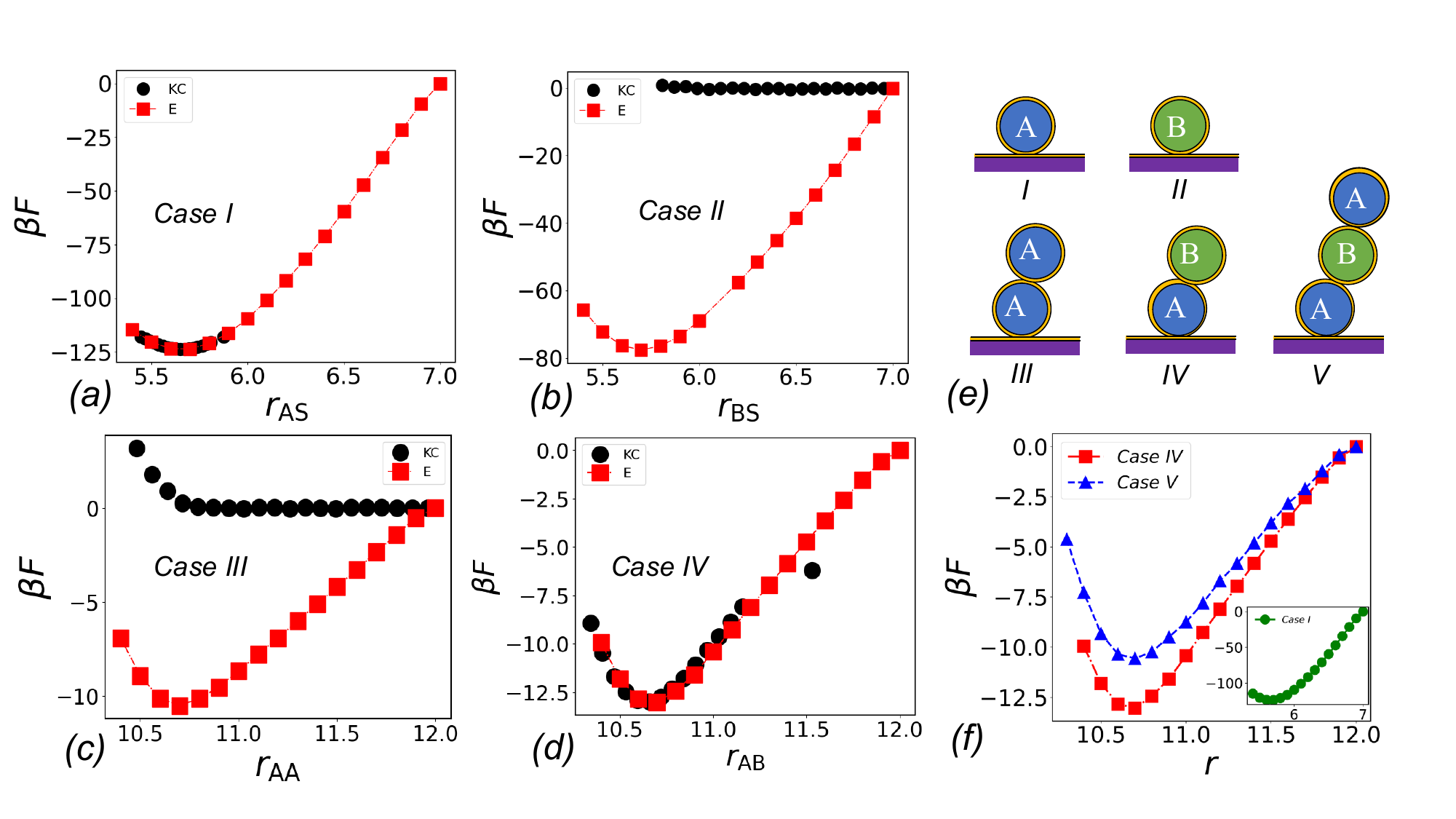}
    \caption{
    {\bf  Kinetically-Constrained (KC) {\em vs} Equilibrium (E) Effective Potentials.} KC (black circles) and E (red squares) effective potentials for a single particle near the surface (a, b), and a particle near another A-type particle bound to the surface (c, d).  In simulations,
    A-type particles bind the surface (a) but do not bind surface-bound A particles (c). Instead, B-type particles bind surface-bound A particles (d) and are repelled by the surface (b). In all cases, E effective potentials are attractive. 
     (e) Configurations employed in the calculation of the effective potentials. In III-V the surface-bound colloid is fixed as the B-type colloid in V. The effective potentials are then calculated as a function of the distance between the particle and the surface (I and II) or between the two particles at the top (III-V). (f) E effective potential of a particle binding the surface (a), a surface-bound particle (d), and two bound colloids (V in panel e). The calculation of E effective potentials are detailed in SI sec.~S5.\cite{di2016communication} We used $N_L=50$, $\rho_S = 2\, L^{-2}$, $\beta\Delta G_0=-10$, and $\beta\Delta G_T=-2$. }
    \label{Fig:potential}
\end{figure}

\section{Results}

\subsection{Two- and Three-Particle Systems}\label{Sec:few_particles}

In this section, we study the effective interaction between colloids and clusters of colloids at the surface. In doing so, we assess the ability of the proposed system to direct the early stages of the self-assembly process towards configurations favoring the targeted structure. 
The comparison between reaction-diffusion simulations\cite{Petitzon_SoftMatt_2016,jana2019surface} (Sec.~\ref{Sec:Simulation}) and free energy calculations\cite{di2016communication} (SI Sec.~S5) clarifies how engineering reaction kinetics can constrain the self-assembly equilibrium landscape  when starting from diluted colloidal suspensions.

In Figs.\ \ref{Fig:potential}$a$ and $b$, we consider single particles at the surface and calculate the  kinetically-constrained (KC) effective interaction
(black circles) using Boltzmann inversion of the particle-surface distance histograms, $p(r)$,
\begin{eqnarray}
\beta F=-\log\left[p(r)/r^2\right]+\log\left[p(r_\mathrm{max})/r_\mathrm{max}^2\right]\quad ,
\label{Eq:BI}
\end{eqnarray}
where $r_\mathrm{max}$ is the maximum particle-plane distance allowed.  In Fig.~\ref{Fig:potential}, the kinetic simulations did not sample the entire attractive part of the potential, so we fit the second term of Eq.~\ref{Eq:BI} to maximize overlap with simulations.
We consider initial configurations in which the particle is self-protected and forms the maximum number of possible loops ($N_L$). 
A-type particles are firmly attracted to the plane (circles in Fig.\ \ref{Fig:potential}$a$) while B-type particles only sense repulsive interactions (Fig.\ \ref{Fig:potential}$b$). This is mirrored by the formation of toehold-mediated bridges between the A particle (but not B) and the surface. When comparing  KC 
with  equilibrium (E) effective potentials (red squares), the latter calculated using multivalent free-energies\cite{di2016communication} (see SI Sec.~S5), we find an agreement 
only for A particles.  (For simplicity, in Fig.~\ref{Fig:potential} both potentials are labeled by $F$.)

The results of Figs.\ \ref{Fig:potential}$a$ and $b$ are expected to guide the self-assembly process towards the formation of a first layer made of  A-type particles. To assess the formation of the second layer, in Figs.\ \ref{Fig:potential}$b$ and $c$, we study the effective potentials between a particle (respectively, of type A and B) and a surface-bound colloid A  (Case III and IV in Fig.~\ref{Fig:potential}$e$). In our simulation, the latter particle is fixed (at a distance $R+L$ from the surface) while the second colloid is free to move.   
Using this setting, we find that,  in simulations, A-type particles (but not B-type particles) are repelled from the surface-bound colloid despite an attractive  equilibrium effective potential. 
 In general, we expect some disagreement between KC and E potentials also for Case IV in Fig.~\ref{Fig:potential}$e$ due to the fact that equilibrium calculations consider three-strand complexes which are not kinetically accessible (Fig.~\ref{Fig:Design}$c$ and SI Fig.~1). Such discrepancies are not visible in Fig.~\ref{Fig:potential}$d$ because we fit the second term of Eq.~\ref{Eq:BI} (see above).
 It is important to stress that the KC effective potentials of Fig.~\ref{Fig:potential} are history dependent. Starting configurations featuring free linkers would lead to attractive KC effective potentials.

 In Fig.~\ref{Fig:potential}$f$, we compare the E effective potentials for Case I, IV, and V (Fig.~\ref{Fig:potential}$e$). Such effective potentials quantify the attraction exerted by a crystallite (with 0, 1, and 2 layers, respectively) on a particle in the fluid phase. The possibility of forming aggregates with finite thickness arises from the fact that this attraction decreases with the number of layers.\cite{jana2019surface,jana2020self}

 Overall, Fig.\ \ref{Fig:potential} shows that effective interactions calculated from kinetic simulations are  either negligible or consistent with those derived from free energies.  Therefore, toehold-exchange designs can act as a digital filter, selectively suppressing some of the equilibrium  particle-particle interactions.

\begin{figure}
    \centering
    \includegraphics[scale=0.6,trim=5.5cm 4cm 0cm 2cm,clip]{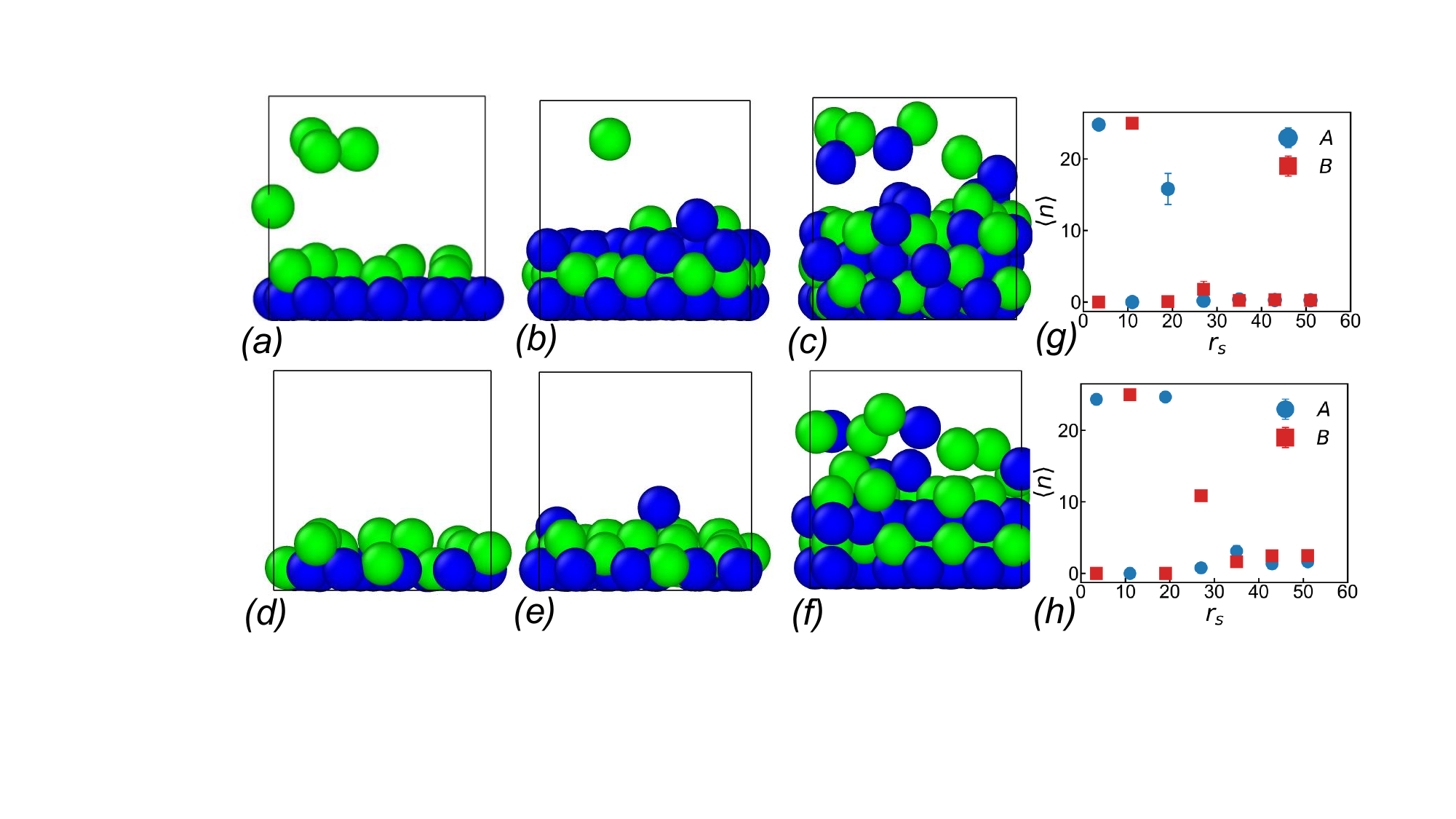}
    \caption{{\bf Colloidal layer deposition with a controllable number of layers and compositional order.} 
    Multi-particle reaction-diffusion simulations result in the self-assembly of crystals made of a controllable number of stacked, same-type colloid planes. The number of planes is controlled by $N_L$ ($N_L=50$, 75, and 100 in a, b, and c) and the chemical potential ($\rho_{\it id}=10^{-7}$, $10^{-6}$, and $10^{-4}$ in d, e, and f). In the top and bottom row we set $\rho_{\it id}=10^{-5}$ and $N_L=75$, respectively. In all panels we set $\rho_S=0.28\, L^{-2}$, $\beta\Delta G_0=-20$, and $\beta\Delta G_T=-2$.  (g,h) Number of A-type and B-type particles found at different layers (identified by the particle-surface distance $r_S$) for system parameters corresponding to panel b and f, respectively. The error bars were calculated by averaging over 
    independent simulations. 
    }
    \label{Fig:many-particle_rendering_1}
\end{figure}

\subsection{Multi-Particle Systems}\label{Sec:many_particles}

Multi-particle simulations are implemented using the grand-canonical (GC) ensemble. These simulations are controlled by the ideal concentration $\rho_{\it id}$ of an ideal gas with chemical potential ($\mu$) equal to the chemical potential of the colloids.\cite{frenkel2023understanding}\footnote{In particular, $\mu=\mu_A=\mu_B\approx k_B T\log \rho_{\it id}$, if $\mu_A$ and $\mu_B$ are, respectively, the chemical potential of colloids of type A and B.} The GC algorithm attempts to remove only particles that do not form any bridge with other particles or the surface. Accordingly, the insertion move adds self-protected particles featuring only loops and no bridges.

Each step of the simulation comprises a reaction dynamics step, a Brownian update, and, with a given probability, a grand canonical move.  We start from empty simulation boxes and let the system evolve until a steady configuration is reached. SI Figs.\ S2 and S3 report the number of particles in the system {\em versus} time for the system parameters explored in Fig.\ \ref{Fig:many-particle_rendering_1}.

Figs.\ \ref{Fig:many-particle_rendering_1}$a$-$f$ report typical steady configurations. The simulations yield expected colloidal structures comprising stacks of layers of the same type of colloids. 
 In Figs.~\ref{Fig:many-particle_rendering_1}$g$-$h$ and SI Fig.~4, we assess the compositional order of the aggregates by reporting the average number of particles of colloids of type A and B as a function of the distance from the surface ($r_S$), binned into intervals that overlap with the position of the expected crystal layers. The irreversibly adsorbed layers are free of defects, except for the systems with the highest value of $N_L$ (SI Fig.~S4$a$ and Fig.~\ref{Fig:many-particle_rendering_1}$c$).
Similarly to what was found in our previous study,\cite{jana2019surface,jana2020self,lanfranco2020adaptable} Fig.\ \ref{Fig:many-particle_rendering_1} shows how the number of layers is controlled by the number of ligands {\em per} particle (Fig.\ \ref{Fig:many-particle_rendering_1}$a$-$c$) and the ideal concentration controlling the bulk particle density (Fig.\ \ref{Fig:many-particle_rendering_1}$d$-$f$). In particular, the number of layers tends to increase with the density of the colloid and the number of DNA strands {\em per} particle. Increasing the number of DNA linkers is a way of increasing particle--particle and particle--surface interactions through combinatorial entropy.\cite{mognetti2019programmable}  Theory results (see SI Sec.~S5) supporting the findings of Fig.~\ref{Fig:many-particle_rendering_1}$a$-$f$ are reported in SI Fig.~S5.

\begin{figure}
    \centering
   \includegraphics[scale=0.55,trim=1.2cm 7cm 0cm 1cm,clip]{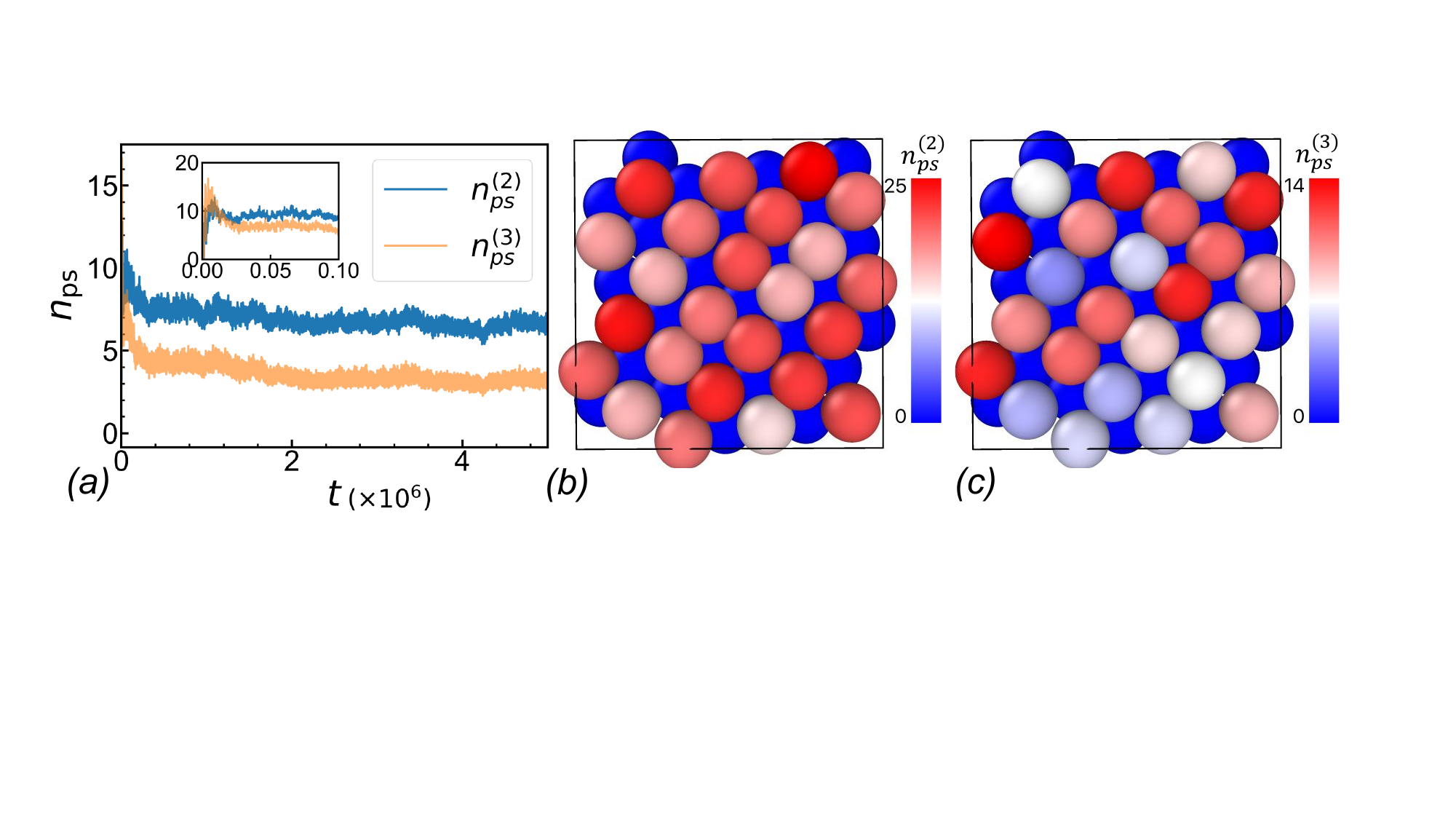}
    \caption{{\bf Evolution of the number of complexes.} (a) Average number of two- ($b_1a_1$ bridges) and three- ($a_1a_2b_1$) strand complexes ($n^{(2)}_{\it ps}$ and $n^{(3)}_{\it ps}$, respectively), formed between A-type colloids and the surface as a function of time. (b,c) Bottom view of the steady configurations. The color map highlights the number of (b) two- and (c) three-strand complexes featured by colloids in contact with the surface. 
    We set $N_L=75$, $\rho_S=0.28 \, L^{-2}$, $\beta \Delta G_0 = -20$, $\beta \Delta G_T = -2$, and $\rho_{\it id}=10^{-5}$.
}\label{Fig:complexes_visualization}
\end{figure}

Fig.\ \ref{Fig:complexes_visualization} studies the number of two- and three-strand complexes that graft the particles in the first layer to the surface. Fig.\ \ref{Fig:complexes_visualization}$a$ shows how such numbers of complexes relax towards their steady value within timescales that are comparable (at sufficiently high $\rho_{\it id}$) with the relaxation of the number of particles in the system (SI Figs.\ S2 and S3). This implies that the assembly process is limited by reaction events (as discussed below). In Fig.\ \ref{Fig:complexes_visualization}$b$ and $c$, we use a color map to visualize, respectively, the two- and three-strand complexes (respectively, $n^{(2)}_{\it ps}$ and $n^{(3)}_{\it ps}$) under steady conditions. The typical number of three-strand complexes featured by the colloids is smaller than that of the two-strand complexes. Three-strand complexes are designed to be transient configurations, enabling the transition between bridges and loops (Fig. \ref{Fig:Design}) without significantly impacting thermodynamics.\cite{parolini2016controlling} 
Figs.\ \ref{Fig:complexes_visualization}$b$ and $c$ show the presence of large fluctuations in the number of complexes. Such fluctuations are due to the stochastic reaction dynamics employed in the simulation and are magnified by configurational disorder and the presence of defects.


\begin{figure}
    \centering
\includegraphics[scale=0.55,trim=1.5cm 7cm 0cm 1cm,clip]{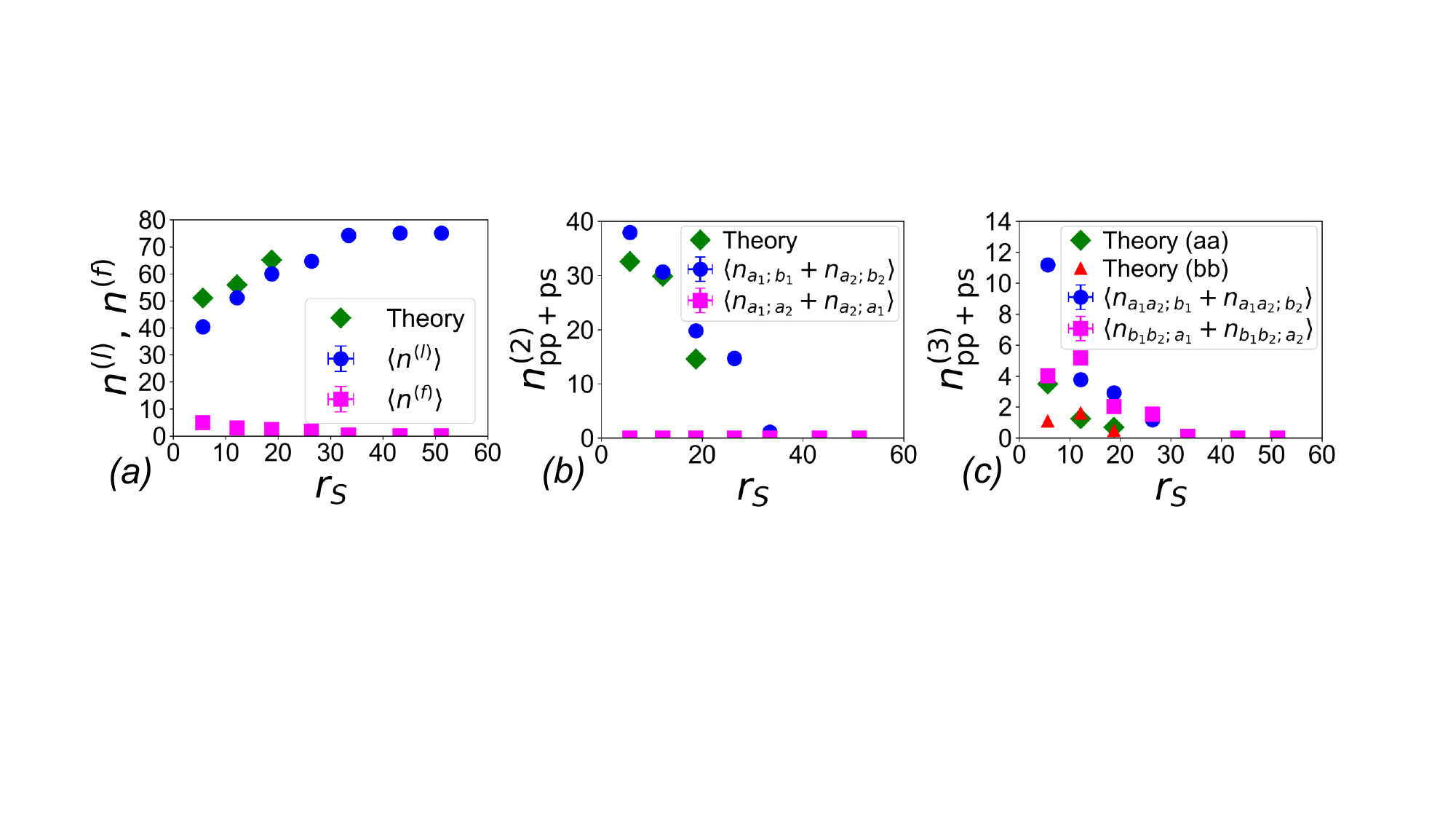}
\caption{{\bf Number of complexes found at different layers.} The figure reports the average number of different complexes featured by colloids belonging to a given layer identified by the colloid-surface distance ($r_S$). (a) Number of loops ($n^l$, circles) and free strand ($n^f$, squares). (b) Two-strand bridges formed by $a$ and $b$ ligands (circles), and by two $a$ ligands (squares). (c) Three-strand complexes featuring two $a$ ligands (circles) and two $b$ ligands (squares). Theoretical predictions are reported with diamond  and triangular symbols. We set $\rho_S=0.28\, L^{-2}$, $N_L=75$, $\beta\Delta G_0=-20$, $\beta\Delta G_T=-2$, and $\rho_{\it id}=10^{-5}$.  
    }
    \label{Fig:complexes}
\end{figure}

Fig.\ \ref{Fig:complexes} extends the analysis of Fig.\ \ref{Fig:complexes_visualization} by reporting the average number of different complexes found on the particles belonging to a given layer, the latter identified by the average particle-surface distance ($r_S$).
We consider a system that yields aggregates made of three stable layers (Fig.~\ref{Fig:many-particle_rendering_1}$b$). In particular, the data points with $r_S>20$ refer to bulk particles or particles that transiently bind the aggregate.   
The number of loops increases with $r_S$ (Fig. \ref{Fig:complexes}$a$) until it reaches its maximum possible value for particles in the bulk gas phase.

Fig.\ \ref{Fig:complexes}$b$ studies the number of inter-particle bridges. Blue circles and magenta squares are, respectively, inter- and intra-layer bridges. The latter are absent, given that aggregates are bipartite BCC structures. Notice that, as in our previous contribution,\cite{jana2019surface} particles at the same layer express an excess of DNA linkers of the same type, resulting in weak lateral particle-particle interactions (Fig.\ \ref{Fig:Design}). Because of this result, particles in the second layer are pivotal in ordering colloids in the first layer.

Fig.\ \ref{Fig:complexes}$c$ reports the number of three-strand complexes. A-type particles at the surface form approximately ten three-strand complexes, mainly of type $a_1 a_2 b_1$ with the surface. The number of three-strand complexes decreases when increasing $r_S$. This is because, in bulk, the ground state of ordered structures would feature only two-strand complexes (either loops or bridges). Indeed, the formation of a pair of three-strand complexes from three two-strand complexes requires an input free energy equal to $\Delta G_T-\Delta G_0 > 0$. Such a gap is large in the limit taken by this work, $\Delta G_0 \ll \Delta G_T$. The number of three-strand complexes is controlled by $\Delta G_T$, as shown in SI Figs.\ S6 and S7, which replicate Fig.\ \ref{Fig:complexes}, respectively, for $\Delta G_T=0$ and $\Delta G_T=-4 k_B T$. The number of three-strand bridges is correlated with the number of free strands (Fig.\ \ref{Fig:complexes}$a$). Free strands are only present in the first few layers because of the opening of loops triggered by surface receptors. 
However, in the current design, free bridges can be depleted by the formation of three-strand complexes. Fig.\ \ref{Fig:complexes}$a$ and SI Figs.\ S6 and S7 show that this is the case when decreasing $\Delta G_T$.

The timescale taken by the system to reach the steady state is not affected by $\Delta G_T$ (SI Fig.~S8). This can be understood by the fact that the rate at which three-strand complexes are formed is not a function of $\Delta G_T$ (Eq.\ \ref{Eq:rates_3}). Once a three-strand complex is formed, it can transform into a bridge or a loop plus a free linker with equal probability, and such a transition is faster when increasing $\Delta G_T$ (Eq.\ \ref{Eq:rates_3}). This is because we neglect the time scales at which the central module of a sticky end ($\alpha$ or $\overline \alpha$) is displaced by an invading DNA strand (second transition in Fig.\ \ref{Fig:Design}$d$).\cite{srinivas2013biophysics} Relaxing the latter assumption would slow down the relaxation process.

In Fig.\ \ref{Fig:complexes} we compare simulation results and theoretical predictions. We prepare a perfect crystal made of three stacked layers and calculate the equilibrium number of complexes featured by such a structure (Eqs.~\ref{Eq:chem_eq}). Theory predictions are in semi-quantitative agreement with simulations. Discrepancies are likely due to crystal defects in simulations and particles that transiently bind the aggregate. Indeed, it should be noted that the binding of a particle triggers changes in the number of complexes of all particles in the aggregate. This mechanism is similar to systems featuring cascade reactions, which have been previously reported in chemical networks.\cite{maslov2007propagation}

\subsection{Kinetic Limitations to Colloidal  Aggregation}\label{Sec:KL}

We can use theory to predict the expected number of layers. 
This is done by estimating the multivalent free energy of the system, $F(n_L)$, as a function of the number of layers ($n_L$). 
Using $F(n_L)$, along with a cell model to estimate the configurational entropic cost of forming an aggregate,\cite{jana2019surface} we can predict the expected number of layers as a function of the chemical potential, $\rho_{\it id}$, and number of linkers $N_L$  (SI Sec.~5).

Using this approach, we find that, in general,  higher chemical potentials than those predicted by theory are required to yield aggregates with a target number of layers  (SI Fig.~S5). This observation points to the fact that aggregation may be reaction-limited. 
This observation is consistent with existing literature. Specifically, Ref.\cite{bachmann2016melting} considered a binary suspension of colloids coated with complementary DNA sticky ends and showed how, in the presence of strong bonds, the aggregation process is arrested. This counterintuitive result is because the attachment of new particles to existing clusters is limited by the slow rate at which existing interparticle bonds break in favor of new bonds. A similar argument has been used to rationalize aggregation of finite-sized condensates made by biopolymers.\cite{ranganathan2020dynamic} 

\begin{figure}
    \centering
    \includegraphics[scale=0.5]{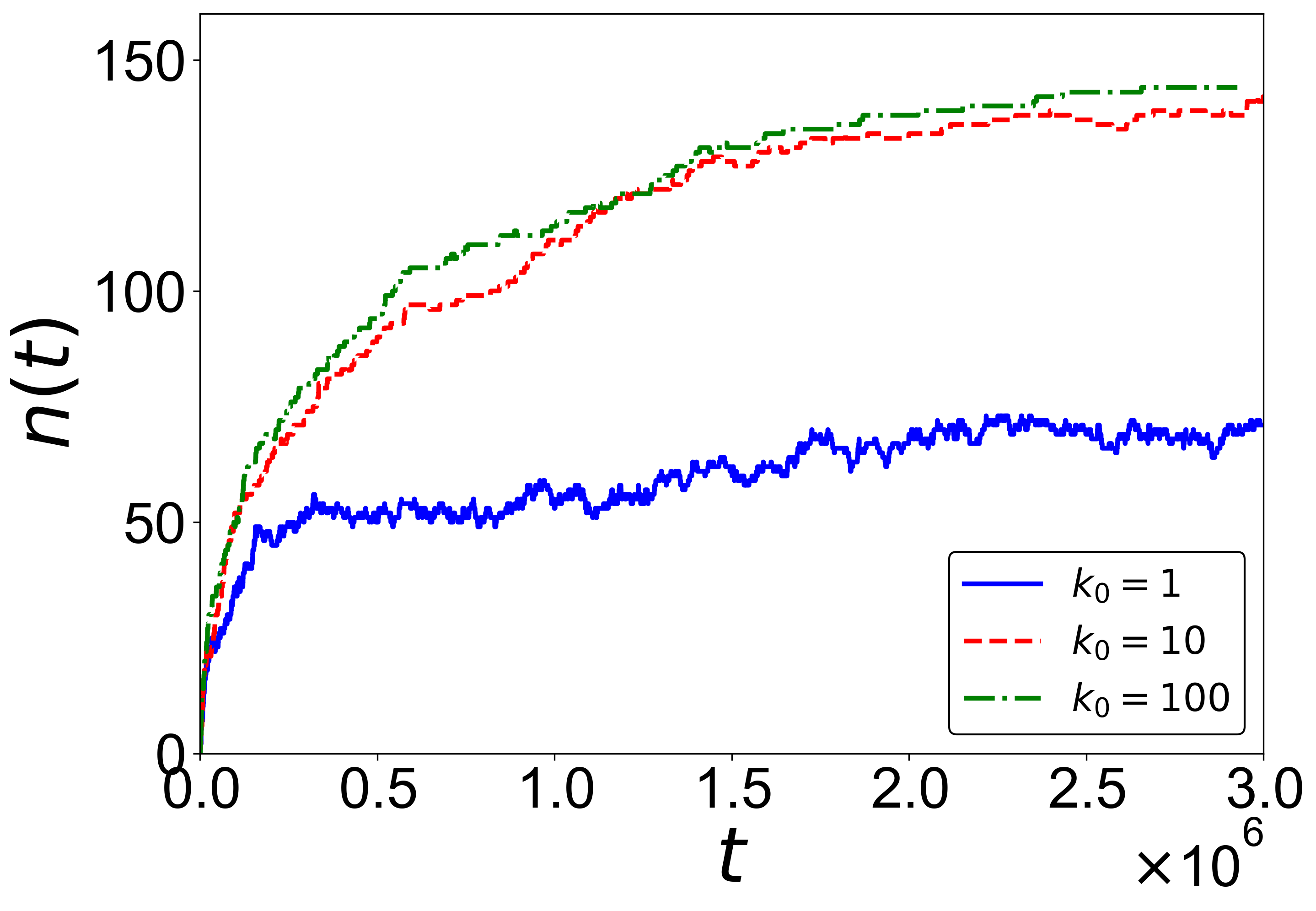}
    \caption{{\bf Self-assembly is limited by reaction kinetics.}  We report the number of particles in the aggregate as a function of time for different values of $k_0$ (Eqs.~\ref{Eq:rates_2}, \ref{Eq:rates_3}). This parameter controls the reaction timescales without affecting the equilibrium states. Increasing $k_0$ leads to larger aggregates. We set $\rho_S=0.28\, L^{-2}$, $N_L=75$, $\beta\Delta G_0=-20$, $\beta\Delta G_T=-2$, and $\rho_{\it id}=10^{-5}$. 
    }
    \label{Fig:kinetics}
\end{figure}
To check if the previous considerations apply to the current system, in Fig.\ \ref{Fig:kinetics} we repeat multi-particle simulations in which we change the factor controlling the absolute value of both the {\it on} and {\it off} reaction rates ($k_0$, see Eqs.~\ref{Eq:rates_2} and \ref{Eq:rates_3}). Given that all chemical equilibrium constants do not depend on $k_0$,  the latter does not alter the equilibrium states of the system. Nevertheless, Fig.\ \ref{Fig:kinetics} shows that the number of planes featured by steady configurations is a function of $k_0$, confirming that the size of the aggregate is still limited by the reaction process.

\section{Discussion \&  Conclusions}
DNA oligomers (or sticky-ends) are employed in programmable self-assembly to engineer the interactions between single units (e.g., functionalized particles). 
Most of the existing systems are based on equilibrium designs, in which particle interactions are inferred using statistical mechanics. Specifically, similar to what is done in standard coarse-graining procedures, particle interactions are calculated as free energies over the ensemble of all possible configurations of DNA oligomers.\cite{Dreyfus_PRL_2009,mognetti2019programmable}

Two aspects challenge the equilibrium design of self-assembly directed by DNA sticky-ends. 
First, it is well known that kinetic limitations can prevent the system from reaching the equilibrium ground state.\cite{biancaniello2005colloidal,zhou2020programming,hensley2023macroscopic,wang2015crystallization} For instance, the notorious stickiness of DNA-mediated interactions\cite{lowensohn2022sliding,marbach2022nanocaterpillar} slows the relaxation of disordered aggregates into crystalline structures. 
More subtly, in advanced systems, equilibrium designs can lead to unavoidable, spurious (i.e., not-sought) interactions. For instance, when considering self-protected colloids carrying complementary sticky-ends (let's say $a_1$ and $a_2)$, interactions between same-type of particles (mediated by $a_1a_2$ bridges) are unavoidable from an equilibrium perspective in which all possible types of complexes between sticky ends are considered (including $a_1a_2$ bridges).

This study proposes to engineer the reaction kinetics to turn off such undesired interactions. 
We consider a system in which the sticky ends are long enough to prevent the denaturation of paired DNA oligomers. Instead, existing bonds are formed/destroyed through strand-displacement reactions catalyzed by the formation of a three-strand complex (toehold-exchange mechanism\cite{zhang2009control}). By properly designing toeholds on the sticky ends, we selectively enable the formation of the desired types of bonds  while inhibiting the formation of the unwanted one. Importantly, such a desired outcome is conditional on having diluted initial configurations with colloids maximising the number of intra--particle loops.

We applied our design to a system leading to colloidal layer deposition with a controllable number of layers. The ability to control the thickness of the aggregate (number of layers) is based on the use of self-protected colloids,  as shown in previous investigations.\cite{jana2019surface,jana2020self,lanfranco2020adaptable} We consider a binary system and use strand-displacement designs to turn off interactions between particles of the same type. As a result, we yield crystallites with a controllable thickness and compositional order.  The current design yields bipartite crystals. More exotic structures, such as non-bipartite crystals with compositional order, could be achieved by a design that activates interactions between particles of the same type only after their incorporation into the aggregate.

This contribution will inspire new designs in which the kinetics will be used to filter thermodynamic  effective interactions, therefore further expanding the set potentialities of DNA-directed self-assembly.

\section*{Data and Software Availability}

Simulation programs, Input files, analysis scripts,  theory scripts, and instructions on how to obtain the data reported in the paper and SI are available at GitHub: 
\url{https://github.com/Aashi0480/DNA_coated_colloids.git}

\begin{acknowledgement}

AKJ, Aashima, and PKJ
acknowledge the financial support from the Science and Engineering Research Board, Government of India
(Grant No. SRG/2022/000993). Aashima also acknowledges the financial support from the
BITS Pilani, Pilani Campus Institute Fellowship. The simulations were performed on the high-performance computing
cluster of BITS Pilani, Pilani Campus. BMM is supported by a PDR grant of the F.R.S.-FNRS (Grant No. T.0158.21) and in part by grant NSF PHY-2309135 to the Kavli Institute for Theoretical Physics (KITP).

\end{acknowledgement}


\appendix 
\setcounter{figure}{0}
\renewcommand{\thefigure}{S\arabic{figure}}

\section{Supplementary Figures}

\begin{figure}[H]
    \centering
    \includegraphics[width=0.75\linewidth]{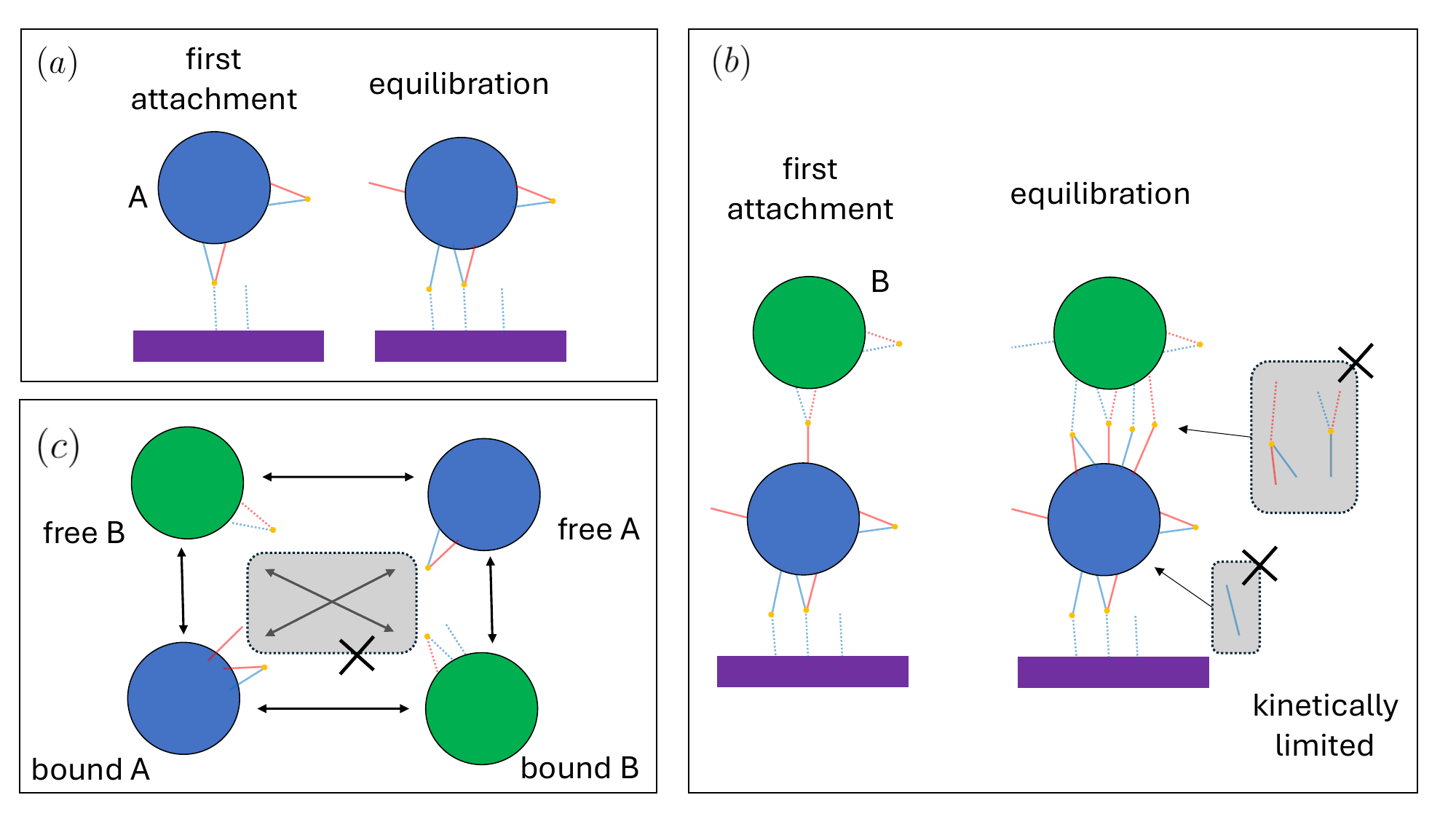}
    \caption{$(a)$ {\bf First layer formation.} Colloids of type A bind the surface through the formation of $a_1a_2 b_1$ complexes. Following the reactions of Main Fig.~1, $a_1a_2 b_1$ complexes generate $a_2$ linkers and $a_2 b_1$ bridges. $(b)$ {\bf Second layer formation.} Binding of colloids of type B to surface-bound A particles is then initiated by the formation of $b_1 b_2 a_2$ complexes. The latter type of complexes generates $a_2 b_2$ bridges and $b_1$ linkers. $b_1$ linkers lead to the formation of $b_1 a_1 a_2$ complexes, $b_1 a_1$ bridges, and $a_2$ linkers on surface-bound A particles. Notice that not all thermodynamically possible three-strand complexes are present and that bound A particles do not express $a_1$ linkers. $(c)$ In general, bound A and B particles feature bridges, loops, and, respectively, $a_2$ and $b_2$ linkers. As a result, colloids of the same type (either bound or free) do not interact.  }
        \label{fig:reaction_pathway}
\end{figure}

\begin{figure}[H]
    \centering
\includegraphics[scale=0.4]{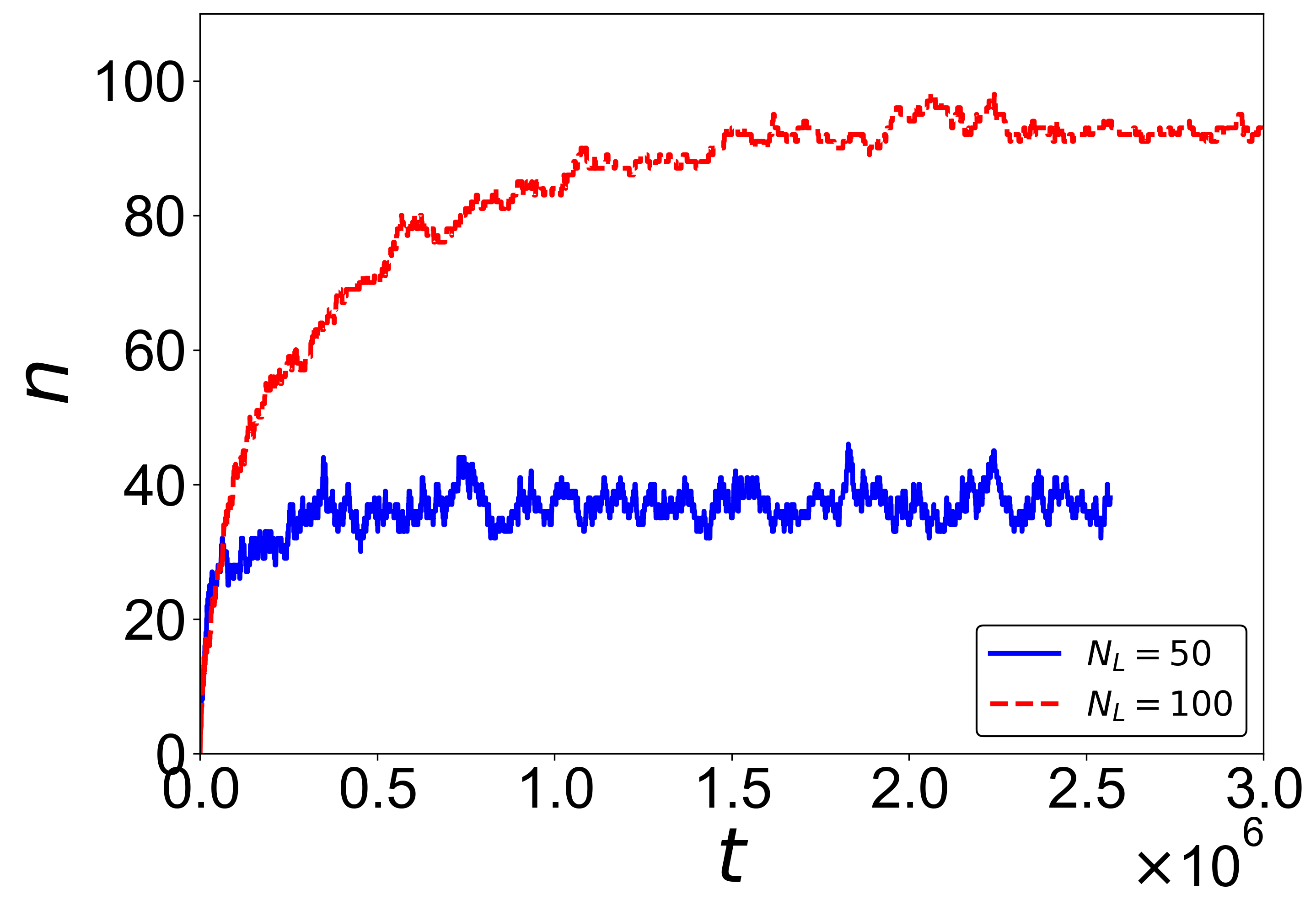}
    \caption{Number of particles {\em vs} simulation time for two different $N_L$. We set $\rho_{\it id}=10^{-5}$, $\rho_S=0.28\, L^{-2}$, $\beta\Delta G_0=-20$, and $\beta\Delta G_T=-2$ (as in Main Fig.~3a and 3c).}
    \label{Fig:particles_time}
\end{figure}

\begin{figure}[H]
    \centering
    \includegraphics[scale=0.4]{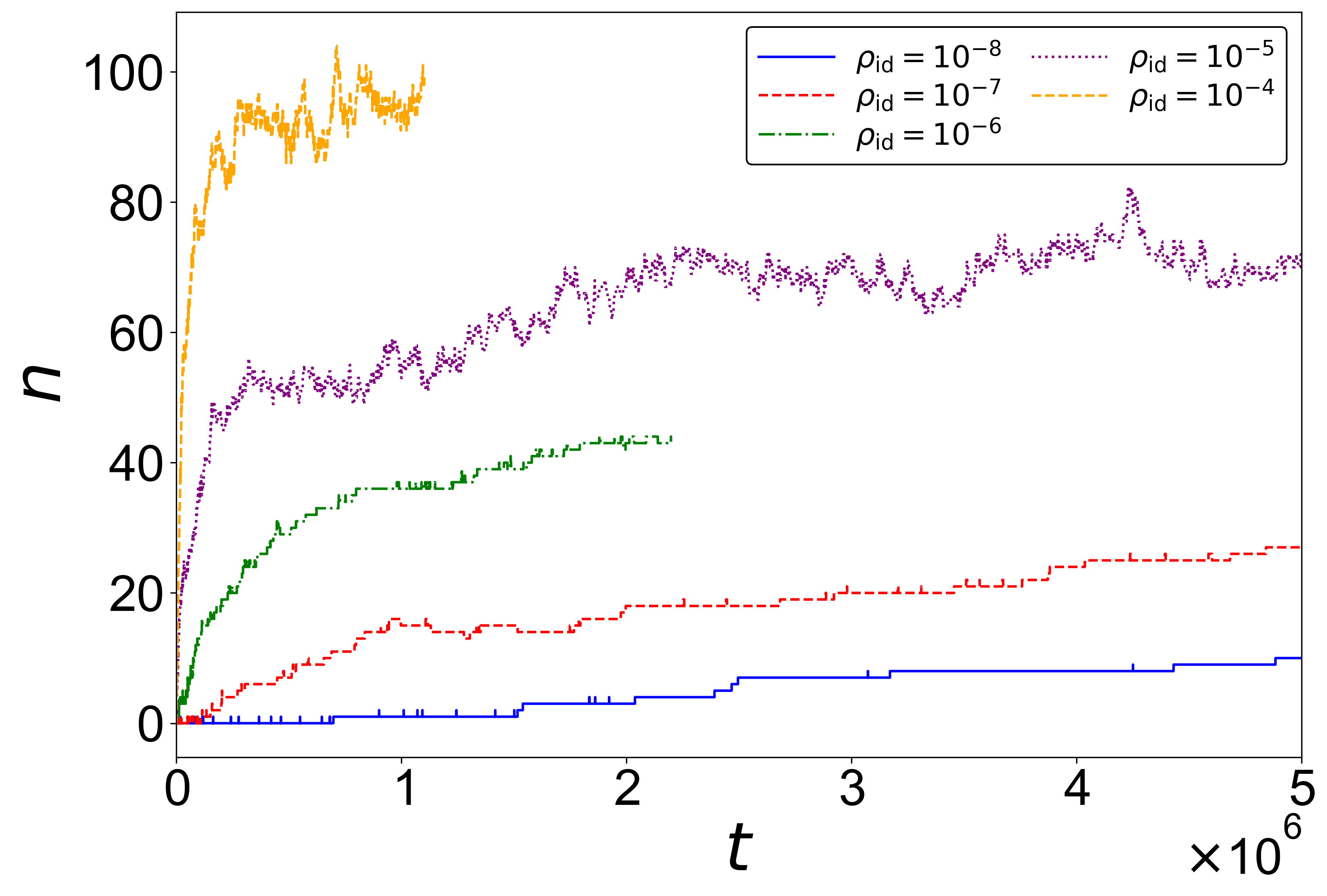}
    \caption{Number of particles {\em vs} simulation time for different chemical potentials $\mu$ ($\mu \sim k_B T \log \rho_{\it id}$). We set $\rho_S=0.28\, L^{-2}$, $\beta\Delta G_0=-20$, and $\beta\Delta G_T=-2$.}
    \label{Fig:particles_time}
\end{figure}

\begin{figure}[H]
        \centering
        \includegraphics[scale=0.55,trim=2.0cm 7cm 0cm 3cm,clip]{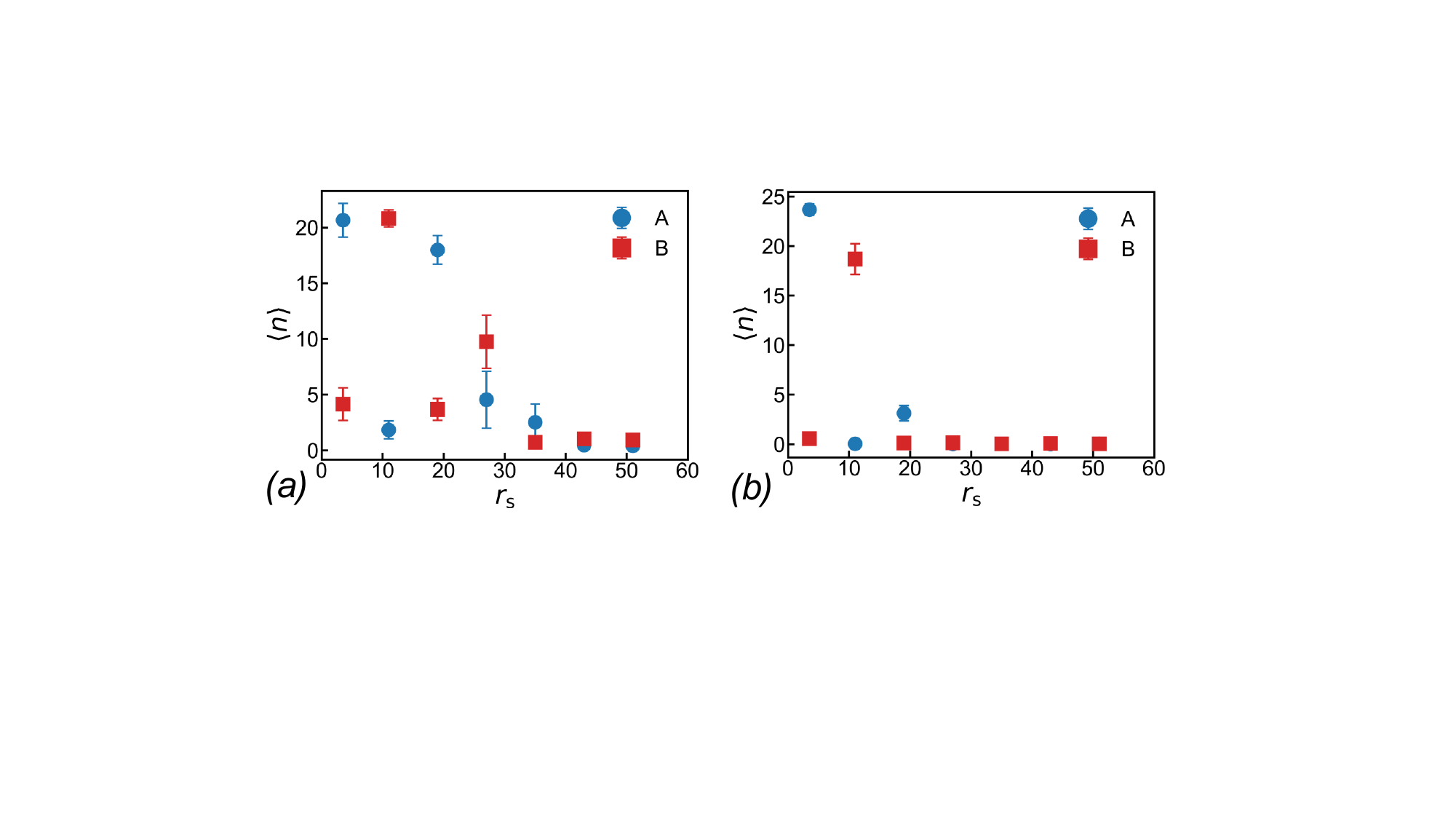} 
    \caption{(a,b)  Same as in Main Fig.~3g for system parameters corresponding to Main Fig.~3c and 3e, respectively. 
    }
    \label{fig:particles_per_layer}
\end{figure}
 
\begin{figure}[H]
    \centering
    \includegraphics[width=0.7\linewidth]{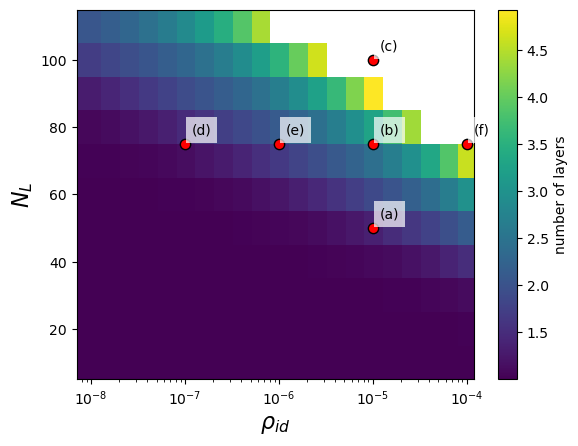}
    \caption{ Theory predictions (based on a cell model, see Sec.~5) of the average number of layers as a function of the number of linkers {\em per} particle ($N_L$) and the chemical potential of the colloids ($\mu \sim \log \rho_{\it id}$). For clarity reasons, we excluded from the color map the top right region (in white), for which the number of layers rapidly diverges. Points correspond to parameters employed in Main Fig.~3. Simulations and theory are in semi-quantitative agreements. Discrepancies are due to approximations employed by the cell model (Sec.~5) and kinetic limitations to the growth of the aggregates (see Main Sec.~3.3).}
    \label{fig:colormap}
\end{figure}

\begin{figure}[H]
    \centering
\includegraphics[scale=0.55,trim=1.5cm 7cm 0cm 3cm,clip]{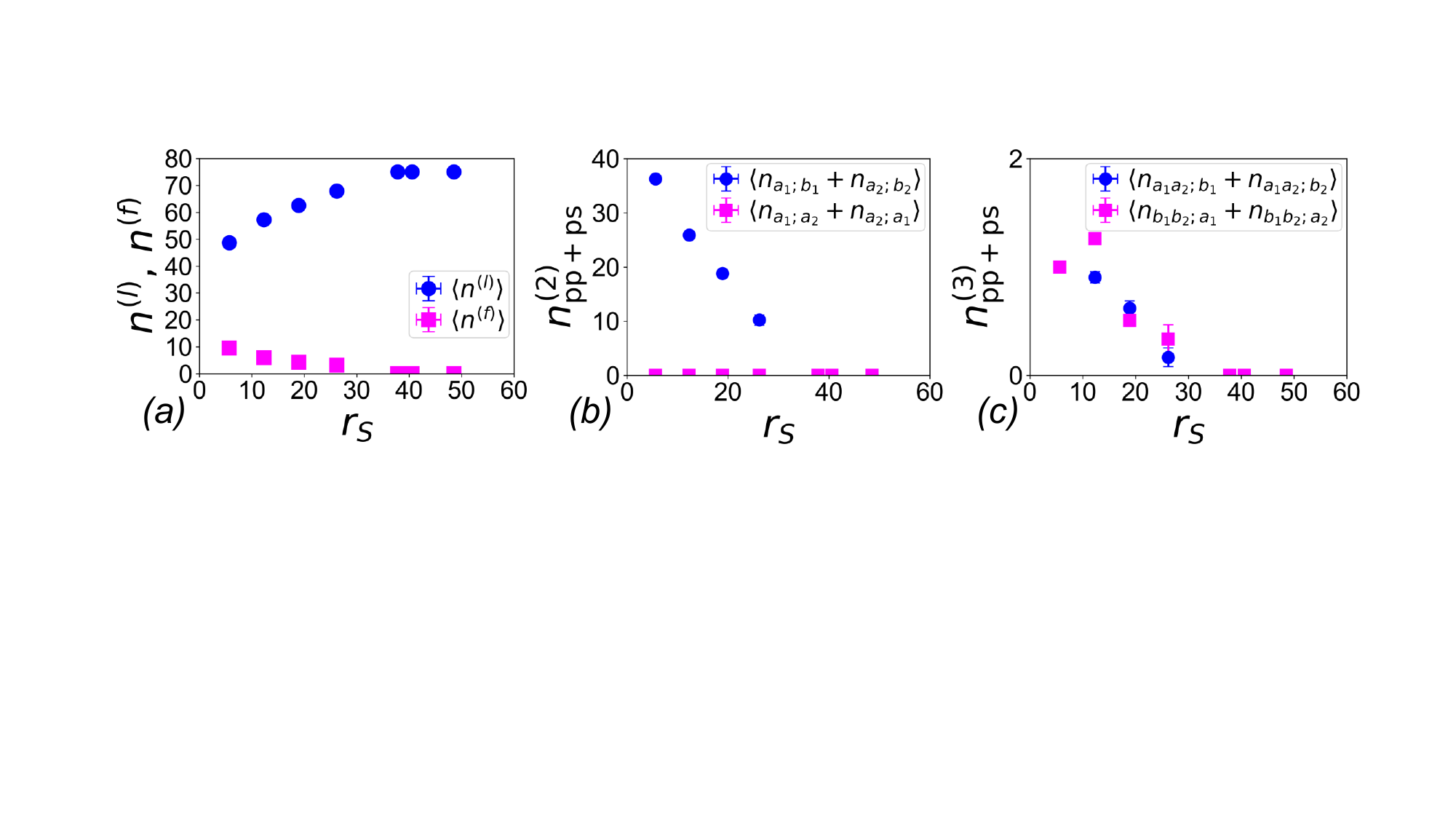}   
    \caption{Same as in Main Fig.~5 but with $\beta \Delta G_T=0$.} 
    \label{Fig:a1b1_l}
\end{figure}

\begin{figure}[H]
    \centering
     \includegraphics[scale=0.55,trim=1.0cm 7cm 0cm 3cm,clip]{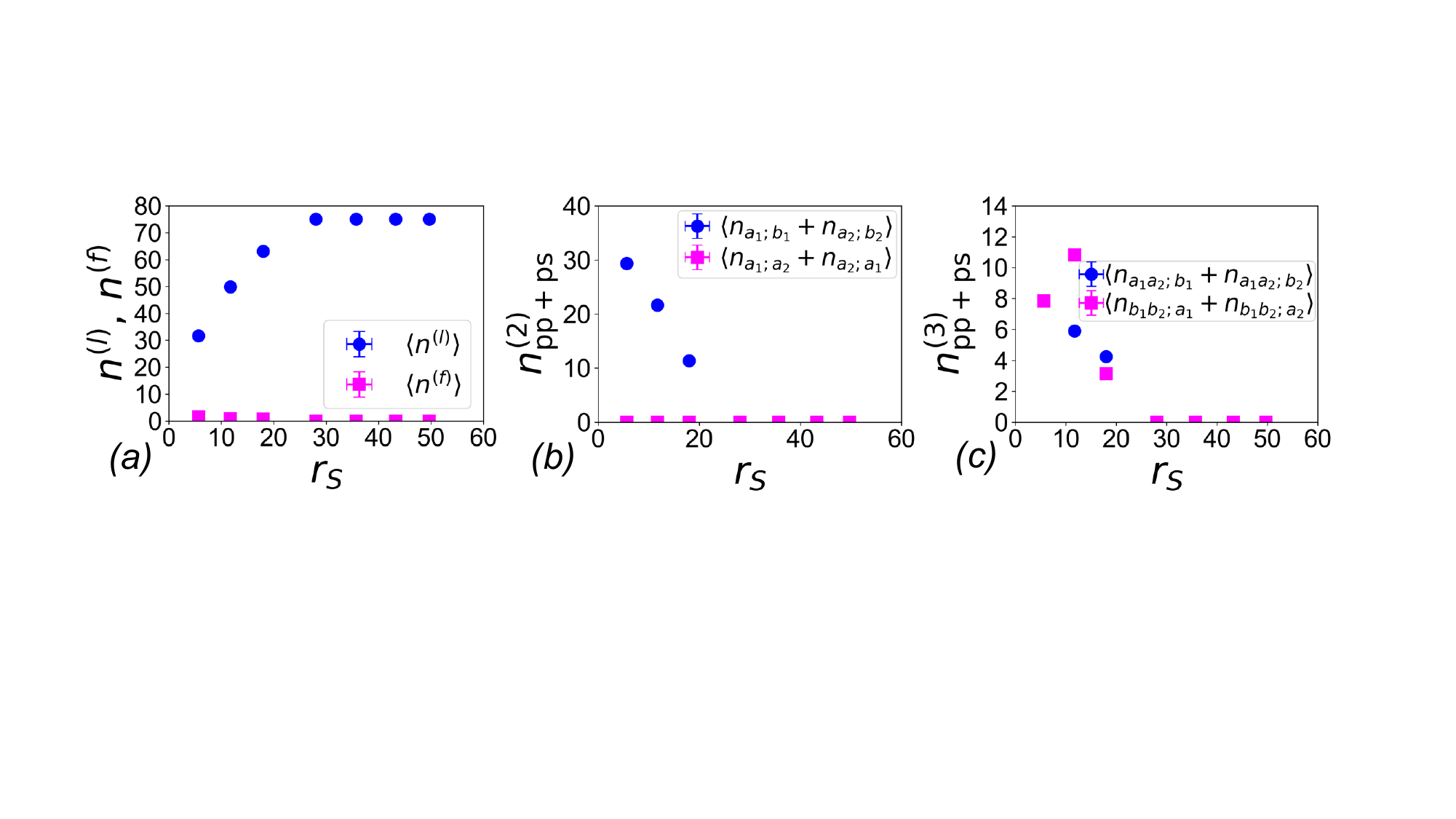} 
    \caption{Same as in Main Fig.~5 but with $\beta \Delta G_T=-4$.}  
    \label{Fig:a1b1_l}
\end{figure}

\begin{figure}[H]
    \centering
    \includegraphics[scale=0.4]{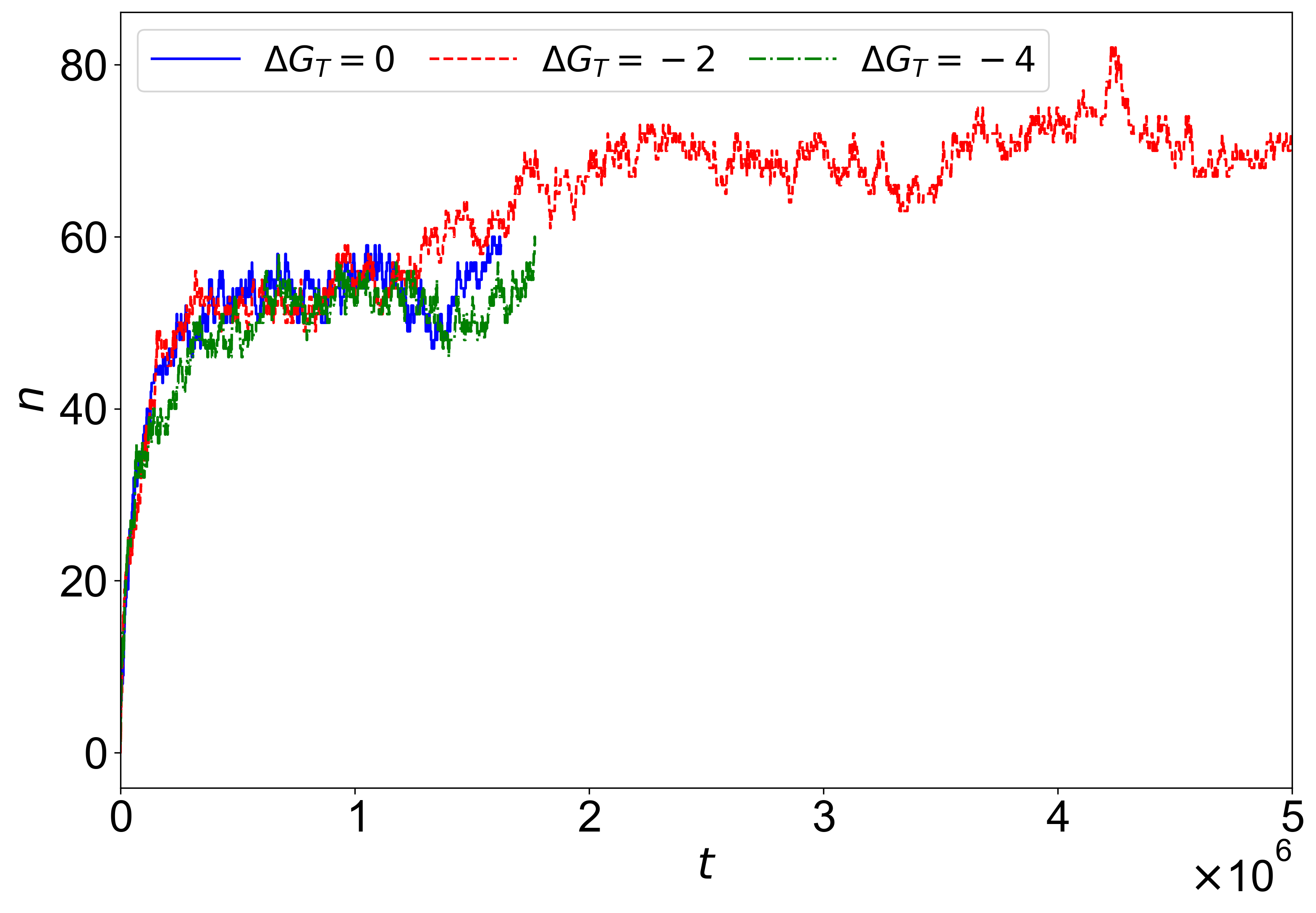}
    \caption{Number of particles {\em vs} simulation time for different $\Delta G_T$. We set $\rho_{\it id}=10^{-5}$, $\rho_S=0.28\, L^{-2}$, $\beta\Delta G_0=-20$, and $N_L=75$.}
    \label{Fig:particles_time}
\end{figure}

\clearpage

\section{Complexes' Configurational Entropies}\label{Sec:Config}

\begin{figure}[h] 
\includegraphics[scale=0.3]{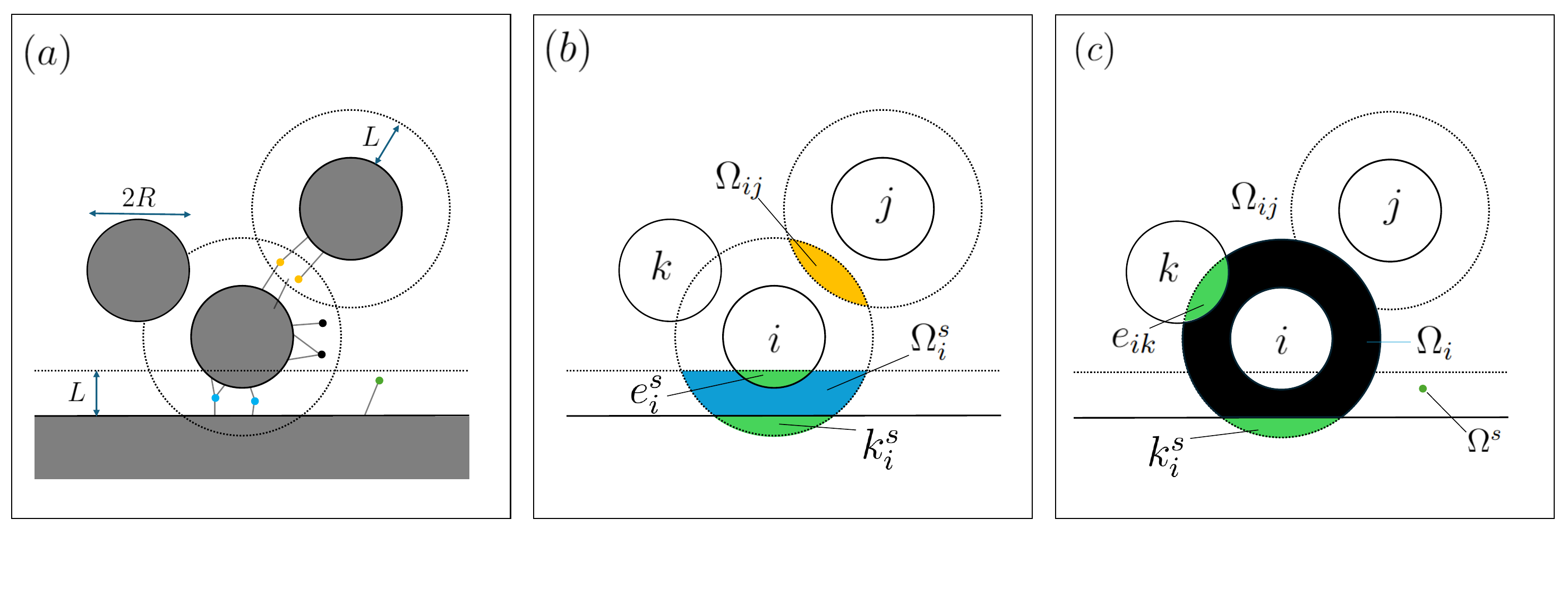} 
\caption{{\bf Complexes' Configurational Entropies.}
$(a)$ The configurational volumes available to complexes play a crucial role in determining the forces between the particles and the reaction equilibrium constants. The configuration volumes of bridges $(b)$ and loops/free strands $(c)$ are calculated as the volume available to the sticky ends.}
\label{Fig:Design_config}
\end{figure}

We assume that the motility of the complexes (ligands, loops, and bridges) is significantly faster than that on any other timescale in the system. Consequently, we do not explicitly track the position and orientation of individual linkers but only the number of complexes of each type present on each colloid and on the surface. The motility of the complexes is incorporated into the model through their configurational volumes (or partition functions). These volumes become analytically tractable under the assumption that non-specific ligand-ligand interactions, such as steric hindrance, are negligible. An additional simplification arises when the ligands are modeled as thin rods with lengths much shorter than the radius of the colloid ($L\ll R$).\cite{Angioletti-Uberti_PRL_2014} Under this condition, the configurational volume of a given complex corresponds to the Euclidean volume accessible to the corresponding point-like sticky ends (Fig.~\ref{Fig:Design_config}).\cite{Angioletti-Uberti_PRL_2014,Petitzon_SoftMatt_2016} The DNA brush of the colloids can then be represented like an ideal gas of reacting/reacted sticky ends uniformly distributed over their accessible regions (Fig.~\ref{Fig:Design_config}$b$ and $c$). The configurational volume of a given complex defines the local concentration of the associated sticky end and therefore affects the reaction dynamics (see Main Sec.~3.1). Configurational volumes also contribute with osmotic terms to the forces between colloids (see Sec.~3.2).

For two- and three-strand bridges, the configurational volume is equal and determined by the intersection of two spheres or a sphere with a plane ($\Omega_{ij}$ and $\Omega^s_i$ in Fig.~\ref{Fig:Design_config}$b$). This volume is eventually reduced by the volumes excluded to the sticky ends by the hard core of the colloids ($e_{si}$ and $e_{is}$ in Fig.~\ref{Fig:Design_config}$b$). Similarly, the configurational volumes of the free linkers and loops moving on particle $i$ and the surface are denoted by $\Omega_i$ and $\Omega^s$, respectively (Fig.~\ref{Fig:Design_config}$c$). The configurational volumes of free strands and loops are reduced by the volumes excluded to the sticky ends by the presence of steric hindrance ($e^s_i$, $e_{ik}$, and $k^s_i$, the latter affecting $\Omega^s$). Detailed expressions for all configurational terms depicted in Fig.~\ref{Fig:Design_config}$b$ and $c$ are reported elsewhere.\cite{jana2019surface,Angioletti-Uberti_PRL_2014}

\section{ Affinities Employed in the Reaction Dynamics Simulations}

Here we list all the affinities ($a$) used in the Gillespie algorithm. The notation follows that used to define the reaction rates (Main Eqs.~2, 3). The index 's' refers to complexes involving the surface. The configurational terms have been defined in SI Sec.~\ref{Sec:Config}.

The affinities relative to the formation/destruction of loops on particles $A$ and $B$ are the following 
\begin{eqnarray}
    a^{\it{on}}_{a_1, a_2, i}  = {k_0  \over \Omega_i\rho_0} n^{a_1}_{i} n^{a_2}_i 
 & \qquad &
a^{\it{off}}_{a_1, a_2, i}  = n^{a_1, a_2}_i  k_0 e^{\beta \Delta G_0}
\\
    a^{\it{on}}_{b_1, b_2, i}  = {k_0  \over \Omega_i\rho_0} n^{b_1}_{i} n^{b_2}_i 
 & \qquad &
a^{\it{off}}_{b_1, b_2, i}  = n^{b_1, b_2}_i  k_0 e^{\beta \Delta G_0}
\end{eqnarray}

The affinities relative to the formation/destruction of two-strand bridges between the same type of particles are the following 
\begin{eqnarray}
a^{\it{on}}_{a_1,i ; a_2, j} =   k_0 {\Omega_{ij} \over \Omega_i \Omega_j \rho_0} n^{a_1}_{i} n^{a_2}_j 
& \qquad &
a^{\it{off}}_{a_1, i; a_2, j} =   n^{a_1; a_2}_{i;j}  k_0  e^{\beta \Delta G_0}
\\
a^{\it{on}}_{b_1,i ; b_2, j} =   k_0 {\Omega_{ij} \over \Omega_i \Omega_j \rho_0} n^{b_1}_{i} n^{b_2}_j 
& \qquad &
a^{\it{off}}_{b_1, i; b_2, j} =   n^{b_1; b_2}_{i;j}  k_0  e^{\beta \Delta G_0}
\end{eqnarray}
Note that, for instance, we do not consider affinities of the type $a^{\it{off}}_{b_2, i; b_1, j}$ as that would double count complexes.

The affinities relative to the formation/destruction of two-strand bridges between different types of particles are the following
\begin{eqnarray}
a^{\it{on}}_{a_1,i ; b_1, j} =   k_0 {\Omega_{ij} \over \Omega_i \Omega_j \rho_0} n^{a_1}_{i} n^{b_1}_j 
& \qquad &
a^{\it{off}}_{a_1, i; b_1, j} =   n^{a_1; b_1}_{i;j}  k_0  e^{\beta \Delta G_0}
\\
a^{\it{on}}_{a_2,i ; b_2, j} =   k_0 {\Omega_{ij} \over \Omega_i \Omega_j \rho_0} n^{a_2}_{i} n^{b_2}_j 
& \qquad &
a^{\it{off}}_{a_2, i; b_2, j} =   n^{a_2; b_2}_{i;j}  k_0  e^{\beta \Delta G_0}
\end{eqnarray}

The affinities relative to the formation/destruction of two-strand bridges between particles and surfaces are the following
\begin{eqnarray}
a^{\it{on}}_{a_1,i;b_1,s}  = k_0 {\Omega^s_i \over \Omega_i \Omega_s \rho_0} n^{a_1}_{i} n^{b_1}_s 
& \qquad & 
a^{\it{off}}_{a_1,i;b_1,s}  = n^{a_1; b_1}_{i,s}  
k_0 e^{\beta \Delta G_0}
\\
a^{\it{on}}_{b_2,i;b_1,s}  = k_0 {\Omega^s_i \over \Omega_i \Omega_s \rho_0} n^{b_2}_{i} n^{b_1}_s 
& \qquad & 
a^{\it{off}}_{b_2,i;b_1,s}  = n^{b_2; b_1}_{i,s}  
k_0 e^{\beta \Delta G_0}
\end{eqnarray}

The affinities relative to the formation/destruction of three-strand complexes forming through loops are the following 
\begin{eqnarray}
a^{\it{on},\ell}_{a_1, a_2,i;b_1,j}  = k_0 {\Omega_{ij} \over \Omega_i \Omega_j \rho_0} n^{a_1 , a_2}_{i} n^{b_1}_j 
&\qquad &
a^{\it{off},\ell}_{a_1, a_2,i;b_1,j}  = n^{a_1 , a_2;b_1}_{i,j} \cdot {k_0 \over (n_\alpha+1)} e^{\beta \Delta G_T}
\\
a^{\it{on},\ell}_{a_1, a_2,i;b_2,j}  = k_0 {\Omega_{ij} \over \Omega_i \Omega_j \rho_0} n^{a_1 , a_2}_{i} n^{b_2}_j 
&\qquad &
a^{\it{off},\ell}_{a_1, a_2,i;b_2,j}  = n^{a_1 , a_2;b_2}_{i,j} \cdot {k_0 \over (n_\alpha+1)} e^{\beta \Delta G_T}
\\
a^{\it{on},\ell}_{b_1, b_2,i;a_1,j}  = k_0 {\Omega_{ij} \over \Omega_i \Omega_j \rho_0} n^{b_1 , b_2}_{i} n^{a_1}_j 
&\qquad &
a^{\it{off},\ell}_{b_1, b_2,i;a_1,j}  = n^{b_1 , b_2;a_1}_{i,j} \cdot {k_0 \over (n_\alpha+1)} e^{\beta \Delta G_T}
\\
a^{\it{on},\ell}_{b_1, b_2,i;a_2,j}  = k_0 {\Omega_{ij} \over \Omega_i \Omega_j \rho_0} n^{b_1 , b_2}_{i} n^{a_2}_j 
&\qquad &
a^{\it{off},\ell}_{b_1, b_2,i;a_2,j}  = n^{b_1 , b_2;a_2}_{i,j} \cdot {k_0 \over (n_\alpha+1)} e^{\beta \Delta G_T}
\end{eqnarray}

The affinities relative to the formation/destruction of three-strand complexes forming through bridges are the following 
\begin{eqnarray}
a^{\it{on},b}_{a_1 , a_2,i;b_1,j}  = k_0 {1 \over \Omega_i \rho_0} n^{a_1; b_1}_{i;j} n^{a_2}_i  
& \qquad &
a^{\it{off},b}_{a_1, a_2,i;b_1,j} = n^{a_1, a_2;b_1}_{i;j} \cdot {k_0 \over (n_\alpha+1)} e^{\beta \Delta G_T}
\\
a^{\it{on},b}_{a_1 , a_2,i;b_2,j}  = k_0 {1 \over \Omega_i \rho_0} n^{a_2; b_2}_{i;j} n^{a_1}_i  
& \qquad &
a^{\it{off},b}_{a_1, a_2,i;b_2,j} = n^{a_1, a_2;b_2}_{i;j} \cdot {k_0 \over (n_\alpha+1)} e^{\beta \Delta G_T}
\\
a^{\it{on},b}_{b_1 , b_2,i;a_1,j}  = k_0 {1 \over \Omega_i \rho_0} n^{a_1; b_1}_{j;i} n^{b_2}_i  
& \qquad &
a^{\it{off},b}_{b_1, b_2,i;a_1,j} = n^{b_1, b_2;a_1}_{i;j} \cdot {k_0 \over (n_\alpha+1)} e^{\beta \Delta G_T}
\\
a^{\it{on},b}_{b_1 , b_2,i;a_2,j}  = k_0 {1 \over \Omega_i \rho_0} n^{a_2; b_2}_{j;i} n^{b_1}_i  
& \qquad &
a^{\it{off},b}_{b_1, b_2,i;a_2,j} = n^{b_1, b_2;a_2}_{i;j} \cdot {k_0 \over (n_\alpha+1)} e^{\beta \Delta G_T}
\end{eqnarray}

The affinities relative to the formation/destruction of three-strand bridges between A particles and the surface are the following 
\begin{eqnarray}
a^{\it{on},\ell}_{a_1, a_2,i; b_1, s}  = k_0 {\Omega^s_{i} \over \Omega_i \Omega_S \rho_0} n^{a_1, a_2}_{i} n^{b_1}_s 
& \qquad& 
a^{\it{off},\ell}_{a_1, a_2, i; b_1, s}  = n^{a_1,a_2; b_1}_{i;s} 
{k_0\over {\color{green} } n_\alpha+1 } e^{\beta \Delta G_T}
\\
a^{\it{on},b}_{a_1, a_2,i; b_1,s}  = k_0 {1 \over \Omega_i \rho_0}  n^{a_1;b_1}_{i;s} n^{a_2}_i 
& \qquad &
a^{\it{off},b}_{a_1, a_2, i; b_1, s}  = n^{a_1 , a_2; b_1}_{i;s} 
{ \over n_\alpha+ 1 }  e^{\beta \Delta G_T}
\end{eqnarray}

\section{Derivation of the Brownian Dynamics equations (Main Eq.~5)}

We follow Ref.~\cite{mognetti2019programmable}. The forces between particles can be derived from the partition function of the system at a given particle positions ($\{{\bf r}_i\}$) and number of complexes ($\{{\bf n}\})$: 
\begin{eqnarray}
    Z(\{n\},\{{\bf r}_i\}) &\approx& \prod_i  (\Omega_i)^{n_i+n_{i,i}} e^{-\beta \Delta G_0 n_{i,i}} \cdot  
    \nonumber \\
    && \qquad \cdot 
    \prod_{i < j} (\Omega_{ij})^{n_{i;j}+n_{2i;j}+n_{i;2j}} e^{-\beta \Delta G_0 n_{i;j} - \beta (\Delta G_0+\Delta G_T) (n_{2i;j}+n_{i;2j})}
    \label{Eq:Z}
\end{eqnarray}
where we extend the definitions given in Main Eq.~5 as follows
\begin{eqnarray}
    n_i &=& \sum_x n^x_i
    \nonumber \\
    n_{i,i} &=& \sum_{x,y} n^{x,y}_i
    \nonumber \\
    n_{i;j} &=& \sum_{x,y} n^{x;y}_{i;j}
    \nonumber \\
    n_{2i;j} &=& \sum_{x,y,z} n^{x,y;z}_{i;j}
    \nonumber \\
    n_{i;2j} &=& \sum_{x,y,z} n^{x;y,z}_{i;j}
    \nonumber
\end{eqnarray}
$\Omega_i$ is the configurational volume of a free ligand on particle $i$, while $\Omega_{ij}$ is the configurational volume of a bridge between particles $i$ and $j$ (SI Sec.~\ref{Sec:Config}). Due to the assumptions of the model (thin, rigid linkers and negligible particle curvature), $\Omega_i$ is also the configurational volume of a loop, and $\Omega_{ij}$ the configurational volume of a three-strand bridge (SI Sec.~\ref {Sec:Config}). In Eq.~\ref{Eq:Z}, we have omitted combinatorial terms given that we work with a fixed number of complexes $\{n\}$. 
It can be shown that the previous expression is equivalent to Eq.~46 of Ref.\cite{mognetti2019programmable}. Using $Z_{\{n\}}$, we calculate the force acting on particle $i$:
\begin{eqnarray}
    {\bf f}_i &=& {\nabla_{{\bf r}_i} Z_{\{n\}} \over Z_{\{n\}}}
    \\
    &=& \sum_{j\in \{i, v(i)\}} (n_j+n_{j,j}) { {\bf \nabla}_{{\bf r}_i} \Omega_j \over \Omega_j } +\sum_{j<p} (n_{j;p}+n_{2j;p}+n_{j;2p}){{\bf \nabla}_{{\bf r}_i} \Omega_{jp} \over \Omega_{jp}} 
\end{eqnarray}
where the first sum is taken over $j=i$ and $j$ cycling over the list of neighboring particles of $i$ ($v(i)$). The expression of ${\bf f}_i$ is further simplified by using
\begin{eqnarray}
\Omega_i = \Omega_0 - \sum_{p \in v(i)} e_{ip}
\end{eqnarray}
where $\Omega_0$ is the configurational space of free ligands moving on an isolated colloid, and $e_{ip}$ depends only on ${\bf r}_i$ and ${\bf r}_p$ (SI Fig.~\ref{Fig:Design_config}). By considering that $\Omega_{jp}$ is only a function of ${\bf r}_j$ and ${\bf r}_p$, we simplify ${\bf f}_i$ as follows:
\begin{eqnarray}
{\bf f}_i &=& \sum_{j\in v(i)} (n_{i;j}+n_{i;2j}+n_{2i;j}) {{\bf \nabla}_{{\bf r}_i} \Omega_{ij} \over \Omega_{ij}} - \sum_{j\in v (i)} \left( (n_i+n_{i,i}) {{\bf \nabla}_{{\bf r}_i} e_{ij} \over \Omega_i }  + (n_j+n_{j,j}) {{\bf \nabla}_{{\bf r}_i} e_{ji} \over \Omega_j }  \right)
\nonumber \\
\label{Eq:SI:BD_f}
\end{eqnarray}
which is the expression presented in Main Sec.~3.2 (Main Eq.~5).
We further develop Eq.~\ref{Eq:SI:BD_f} by adapting it to the case in which particle $i$ is of type $A$ and $B$
\begin{eqnarray}
{\bf f}^{A}_i &=& {\bf f}^{A}_{A,i} + {\bf f}^{A}_{B,i} + {\bf f}^{A}_{S,i} -\beta \nabla_i V_{\it reg}
\nonumber \\
{\bf f}^{B}_i &=& {\bf f}^{B}_{A,i} + {\bf f}^{B}_{B,i} + {\bf f}^{B}_{S,i}-\beta \nabla_i V_{\it reg}
\label{Eq:Fdec}
\end{eqnarray}
${\bf f}^{X}_{Y,i}$ is the contribution of particles of type $Y$
($Y=A$, $B$, or $S$) surrounding $i$
Using the notation employed in Main Sec.~3.1, we find:  
\begin{eqnarray}
{\bf f}^{A}_{A,i} &=& \sum_{j\in v(i)} \Big[ (n^{a_1;a_2}_{i;j}+n^{a_2;a_1}_{i;j}) {\nabla_{{\bf r}_i} \Omega_{ij}\over \Omega_{ij}}-(n^{a_1}_{j}+n^{a_2}_{j}+n^{a_1,a_2}_{j}){\nabla_{{\bf r}_i} e_{ji}\over \Omega_j}-(n^{a_1}_{i}+n^{a_2}_{i}+n^{a_1,a_2}_{i}){\nabla_{{\bf r}_i} e_{ij}\over \Omega_i} \Big]
\nonumber 
\\
\\
{\bf f}^{A}_{B,i} &=& \sum_{j\in v(i)} \Big[ (n^{a_1;b_1}_{i;j}+n^{a_2;b_2}_{i;j}+n^{a_1;b_1,b_2}_{i;j}+n^{a_2;b_1,b_2}_{i;j}+n^{a_1,a_2;b_1}_{i;j}+n^{a_1,a_2;b_2}_{i;j}) {\nabla_{{\bf r}_i} \Omega_{ij}\over \Omega_{ij}}
\nonumber \\
&& \qquad -(n^{b_1}_{j}+n^{b_2}_{j}+n^{b_1,b_2}_{j}){\nabla_{{\bf r}_i} e_{ji}\over \Omega_j} -(n^{a_1}_{i}+n^{a_2}_{i}+n^{a_1,a_2}_{i}){\nabla_{{\bf r}_i} e_{ij}\over \Omega_i} \Big]
\\
{\bf f}^{A}_{S,i} &=& (n^{a_1;b_1}_{i;S}+n^{a_1,a_2;b_1}_{i;S}){\nabla_{{\bf r}_i} \Omega^s_{i}\over \Omega^s_{i}} -n^{b_1}_S {\nabla_{{\bf r}_i} k^s_{i}\over \Omega^s} - (n^{a_1}_{i}+n^{a_2}_{i}+n^{a_1,a_2}_{i}){\nabla_{{\bf r}_i} e^s_{i}\over \Omega_{i}}
\\
{\bf f}^{B}_{A,i} &=& \sum_{j\in v(i)} \Big[ (n^{b_1;a_1}_{i;j}+n^{b_2;a_2}_{i;j}+n^{b_1;a_1,a_2}_{i;j}+n^{b_2;a_1,a_2}_{i;j}+n^{b_1,b_2;a_1}_{i;j}+n^{b_1,b_2;a_2}_{i;j}) {\nabla_{{\bf r}_i} \Omega_{ij}\over \Omega_{ij}}
\nonumber \\
&& \qquad -(n^{a_1}_{j}+n^{a_2}_{j}+n^{a_1,a_2}_{j}){\nabla_{{\bf r}_i} e_{ji}\over \Omega_j} 
-(n^{b_1}_{i}+n^{b_2}_{i}+n^{b_1,b_2}_{i}){\nabla_{{\bf r}_i} e_{ij}\over \Omega_i} \Big]
\\
{\bf f}^{B}_{B,i} &=& \sum_{j\in v(i)} \Big[ (n^{b_1;b_2}_{i;j}+n^{b_2;b_1}_{i;j}) {\nabla_{{\bf r}_i} \Omega_{ij}\over \Omega_{ij}}-(n^{b_1}_{j}+n^{b_2}_{j}+n^{b_1,b_2}_{j}){\nabla_{{\bf r}_i} e_{ji}\over \Omega_j}-(n^{b_1}_{i}+n^{b_2}_{i}+n^{b_1,b_2}_{i}){\nabla_{{\bf r}_i} e_{ij}\over \Omega_i} \Big]
\nonumber 
\\
\\
{\bf f}^{B}_{S,i} &=& n^{b_2;b_1}_{i;S}{\nabla_{{\bf r}_i} \Omega^s_{i}\over \Omega^s_{i}} -n^{b_1}_S {\nabla_{{\bf r}_i} k^s_{i}\over \Omega^s} - (n^{b_1}_{i}+n^{b_2}_{i}+n^{b_1,b_2}_{i}){\nabla_{{\bf r}_i} e^s_{i}\over \Omega_{i}}
\end{eqnarray}
In Eqs.~\ref{Eq:Fdec}, $V_{\it reg}$ is a smooth potential employed to regularize hard-core interactions. Following previous contributions,\cite{jana2019surface,Angioletti-Uberti_PRL_2014} we have used ($r_{ij}$ and $r^z_i$ distances, respectively, between particle $i$ and $j$, and between the plane and particle $i$)
\begin{eqnarray}
V_{\it reg}(\{r\}) = \sum_{i < j} V_{pp}(r_{ij}) + \sum_i V_{ps}(r^z_i)
\label{Eq:Vreg}
\end{eqnarray}
with 
\begin{eqnarray}
V_{pp}(r_{ij}) = 
\begin{cases} 
M k_B T \log \left( 1 - \frac{e_{ij}(r_{ij}, L')}{4 \pi R^2 \cdot (0.75 \cdot L)} \right) 
& r_{ij} < 2 \cdot R + 0.75 \cdot L \\
0 & r_{ij} \geq 2 \cdot R + 0.75 \cdot L
\end{cases}
\end{eqnarray}
and
\begin{eqnarray}
V_{ps}(r^z_i) =
\begin{cases}
M \log \left( 1 - \frac{e_{i}^s(r^z_i, L')}{4 \pi R^2 \cdot (0.75 \cdot L)} \right) & r^z_i < R + 0.75 \cdot L \\
0 & r_{i,z} \geq R + 0.75 \cdot L
\end{cases}
\end{eqnarray}
with $L'=0.75\cdot L$ and $M=500$. $V_{\it reg}(\{r\})$ can be interpreted as due to the osmotic repulsion of a brush of $M$ inert linkers {\em per} particle of length $L'$.

\section{ Calculation of the Equilibrium Interactions}

In this section, we clarify the equilibrium calculation of the effective potentials reported in Main Fig.~2. The free energy of the system $F$ can be written as\cite{Varilly_JChemPhys_2012} 
\begin{eqnarray}
    F=\Delta F_{\it att} + V_{\it rep} + V_{\it reg}
    \label{Eq:F}
\end{eqnarray}
where $V_{\it reg}$ has been defined in SI Eq.\ \ref{Eq:Vreg}. $V_{\it rep}$ is the osmotic repulsion generated by the linkers without 2- and 3-strand complexes (i.e., when $\Delta G_0\to \infty$ and $\Delta G_T \to \infty$)
\begin{eqnarray}
    V_{\it rep}(\{ {\bf r}_i \}) &=& \sum_i 2 N_L \log {\Omega_i \over 4 \pi L R^2} + N_S \log {\Omega^s \over A L }
\end{eqnarray}
where $A$ and $N_S$ are, respectively, the area and the total number of linkers on the surface. ($V_{\it rep}$ is similar but not strictly equal to $V_{\it reg}$. Indeed, for simplicity, $V_{\it reg}$ does not account for multi-body terms.) $F_{\it att}$ is an attractive term accounting for the formation of complexes. A portable expression for multimeric complexes has been obtained in Ref.~\cite{di2016communication}, which, adapted to the current system, reads as follows  
\begin{eqnarray}
    F_{\it att} = \sum_{i,x} N_L \log {\langle n^x_i\rangle \over N_L} + \sum_{i,x,y} \langle n^{x,y}_{i} \rangle + \sum_{i<j,x,y} \langle n^{x;y}_{i;j} \rangle + 2 \sum_{i<j,x,y,z} \langle n^{x,y;z}_{i;j} +n^{x,y;z}_{i;j} \rangle 
    \label{Eq:Fatt}
\end{eqnarray}
where the averages are defined in Main Eq.~1. The previous expression is meant to include contributions from the surface ($i=s$ and/or $j=s$). In Eq.~\ref{Eq:F} we subtract a reference contribution from $F_{\it att}$
\begin{eqnarray}
    \Delta F_{\it att} = F_{\it att} - F_{\it att,ref}
\end{eqnarray}
In Main Figs.~2$a$ and $b$, $F_{\it att,ref}$ is the attractive term (Eq.~\ref{Eq:Fatt}) for an isolated particle forming the maximum number of loops (for isolated surfaces $F_{\it att}=0$). 
In Main Figs.~2$c$ and $d$, $F_{\it att,ref}$ is the attractive term (Eq.~\ref{Eq:Fatt}) for the particle-surface system.

Fig.~\ref{fig:colormap} has been obtained using a cell model.\cite{jana2019surface}. Specifically, the probability of having a crystallite with $n$ layers, $P(n)$, is calculated as follow 
\begin{eqnarray}
    P(n) &=& {1\over Z}  \exp [-\beta F_{att} (n)] (\rho_{\it id} v_0)^n \quad ,
\end{eqnarray}
where $F_{att}$ is the multivalent energy (as calculated above) of the crystallite {\em per} particle in contact with the surface, and $Z$ is a normalization factor. $v_0$ is the configurational volume available to a particle in the crystal phase and has been taken equal to $v_0=L^3$. The average number of layers is then calculated as $\sum_n P(n) n$. The bulk fluid-crystal transition manifold is identified by the system parameters for which $Z$ diverges. In the calculation of $F_{\it att}$, we exclude the possibility of forming complexes repressed by the kinetic pathway of Fig.~\ref{fig:reaction_pathway}.


\providecommand{\latin}[1]{#1}
\makeatletter
\providecommand{\doi}
  {\begingroup\let\do\@makeother\dospecials
  \catcode`\{=1 \catcode`\}=2 \doi@aux}
\providecommand{\doi@aux}[1]{\endgroup\texttt{#1}}
\makeatother
\providecommand*\mcitethebibliography{\thebibliography}
\csname @ifundefined\endcsname{endmcitethebibliography}
  {\let\endmcitethebibliography\endthebibliography}{}

\end{document}